\documentclass[usenatbib]{mnras2}
\usepackage{newtxtext,newtxmath}

\usepackage[T1]{fontenc}
\usepackage{ae,aecompl}

\usepackage{graphicx}	
\usepackage{amsmath}	
\usepackage{ulem}

\usepackage{comment}

\newlength{\figsize}
\newlength{\figdblsize}
\usepackage[usenames,dvipsnames]{xcolor}
\definecolor{frompapers}{rgb}{0.8, 0.36, 0.36}
\definecolor{todo}{rgb}{0.95, 0.52, 0.0}
\definecolor{header}{rgb}{0.55,0.57,0.67}
\definecolor{changes}{rgb}{0.0,0.7,0.0}

\usepackage{float}

\usepackage{amsmath}
\newcommand{\sfe}{\mbox{$\epsilon_\mathrm{ff}$}}
\newcommand{\Ssfr}{\mbox{$\Sigma_\mathrm{SFR}$}}
\newcommand{\nophoto}{\mbox{SFE2noPI}}
\newcommand{\zero}{\mbox{SFE0}}
\newcommand{\ptwo}{\mbox{SFE02}}

\newcommand{\two}{\mbox{SFE2}}
\newcommand{\ten}{\mbox{SFE10}}
\newcommand{\twenty}{\mbox{SFE20}}
\newcommand{\fifty}{\mbox{SFE50}}
\newcommand{\jeansmass}{\mbox{$\mathrm{M_J}$}}
\newcommand{\solarmass}{\mbox{$\mathrm{M_\odot}$}}

\title[The challenge of star cluster simulations]{The challenge of simulating the star cluster population of dwarf galaxies with resolved interstellar medium}

\author[J. M. Hislop et al.]{
Jessica M. Hislop$^{1}$\thanks{E-mail: jmhislop@mpa-garching.mpg.de},
Thorsten Naab$^{1}$,
Ulrich P. Steinwandel$^{2}$,
Natalia Lah\'en$^{1}$,
Dimitrios Irodotou$^{3,4}$,
\newauthor
\ Peter H. Johansson$^{4}$,
and Stefanie Walch$^{5}$
\\
\\
$^{1}$Max-Planck-Institute f\"ur Astrophysik, Karl-Schwarzschild-Str 1, D-85748 Garching, Germany\\
$^{2}$Center for Computational Astrophysics, Flatiron Institute, 162 5th Avenue, New York, NY 10010, USA\\
$^{3}$Astronomy Centre, University of Sussex, Falmer, Brighton BN1 9QH, UK\\
$^{4}$Department of Physics, University of Helsinki, Gustaf Hällstr\"omin katu 2, FI-00014, Helsinki, Finland\\
$^{5}$Physikalisches Institut der Universit\"at zu K\"oln, Z\"ulpicher Strasse 77, D-50937 K\"oln, Germany\\
}

\date{Accepted XXX. Received YYY; in original form ZZZ}

\pubyear{2021}

\begin{document}
\label{firstpage}
\pagerange{\pageref{firstpage}--\pageref{lastpage}}
\maketitle

\begin{abstract}

We present results on the star cluster properties from a series of high resolution smoothed particles hydrodynamics (SPH) simulations of isolated dwarf galaxies as part of the \textsc{Griffin} project. The simulations at sub-parsec spatial resolution and a minimum particle mass of 4 \solarmass\ incorporate non-equilibrium heating, cooling and chemistry processes, and realise individual massive stars. The simulations follow feedback channels of massive stars that include the interstellar-radiation field variable in space and time, the radiation input by photo-ionisation and supernova explosions. Varying the star formation efficiency per free-fall time in the range \sfe\ = 0.2 - 50$\%$ neither changes the star formation rates nor the outflow rates. While the environmental densities at star formation change significantly with \sfe, the ambient densities of supernovae are independent of \sfe\ indicating a decoupling of the two processes. At low \sfe, gas is allowed to collapse more before star formation, resulting in more massive, and increasingly more bound star clusters are formed, which are typically not destroyed. With increasing \sfe\ there is a trend for shallower cluster mass functions and the cluster formation efficiency $\Gamma$ for young bound clusters decreases from $50 \%$  to $\sim 1 \%$ showing evidence for cluster disruption. However, none of our simulations form low mass ($ < 10^3$ \solarmass) clusters with structural properties in perfect agreement with observations. Traditional star formation models used in galaxy formation simulations based on local free-fall times might therefore be unable to capture star cluster properties without significant fine-tuning.   

\end{abstract}





\section{Introduction}
\label{sec:Introduction}
Recently there has been significant progress in the optimisation of computer algorithms and in the increased capability of high-performance computing systems. Together with an improved numerical implementation of the physical processes governing the evolution of the galactic interstellar medium (ISM) numerical simulations are able to describe the evolution of entire galaxies including a realistic multi-phase ISM component. This is an important step forward \citep[see e.g.][for a review]{NaabOstriker2017} in the understanding of galaxy evolution, as the galactic ISM is the location of all star formation and most of the metal enrichment in the Universe. In addition, the ISM is the driving site for galactic outflows and the origin of most galactic observables at all cosmic epochs.

These next generation simulations have pushed the use of sub-grid models to ever smaller scales. For some cosmological simulations, entire (>100 pc) patches of the ISM are modelled with sub-resolution models \citep[see e.g.][for a review]{Somerville2015}. However, many new high-resolution galaxy formation simulations can resolve the multi-phase ISM structure down to $\sim$ parsec scale. This allows to partially follow important physical processes setting the ISM properties directly, such as the impact of individual supernova (SN) explosions by the approximation of thermal energy injection. Recent simulations also represent the galactic stellar population with increasingly lower mass stellar tracers down to populations of several thousands \citep[e.g.][]{2018MNRAS.480..800H,2021arXiv210306882K,2019MNRAS.489.4233M} or several tens to several hundreds solar masses \citep[][]{2011MNRAS.417..950H,2015MNRAS.446.2038R,2015MNRAS.451...34R,2017MNRAS.464.3580D,2020MNRAS.491.1656A,2020MNRAS.493.4315M,2021MNRAS.505.1678J}. The highest resolution studies have even started to trace individual massive stars for isolated galaxy models \citep[e.g.][]{Hu2016,Emerick2018,Lahen2020,2021MNRAS.501.5597G,2021MNRAS.506.3882S,2021PASJ...73.1036H}. While individual stars are the lowest possible resolution element, also these simulations still rely on sub-grid models for estimating the star formation rates, for the sampling of individual stars and, to some degree, for the modelling of their radiation, energy and momentum output. In this study, we focus on the star cluster population properties in simulations of entire galaxies which have the potential to resolve the multi-phase ISM structure as well as the internal structure of star clusters.  

The majority of the recent high resolution galaxy formation studies, including those mentioned above assume an underlying simple sub-grid model which estimates the local star formation rate based on the local gas density divided by its free-fall time  multiplied with an efficiency \sfe\ parameter \citep{1959ApJ...129..243S} [see equation (\ref{eq:localSFR}) below]. The star formation efficiency per free-fall time based sub-grid model is the most commonly adopted model in all numerical galaxy formation research \citep[see e.g.][]{NaabOstriker2017} and is used with all major simulation methods, i.e. grid codes \citep[e.g.][]{1999PhDT........25K,2002A&A...385..337T,2014ApJS..211...19B}, moving mesh codes \citep[e.g.][]{2010MNRAS.401..791S}, and particle based hydrodynamics codes \citep[e.g.][]{Springel05,2015MNRAS.450...53H}. The different models have varying additional constraints on the properties of the gas particles which become eligible for star formation in the first place. For simulations with resolved ISM this typically refers to the (collapsing) dense and cold gas phase. 

The success of this star formation sub-grid model is based on observational evidence at all cosmic epochs that the star formation rate scales with the gas surface density and is an inefficient process \citep{1959ApJ...129..243S,1998ApJ...498..541K,Leroy2008, Genzel10, Tacconi13}. For typical galaxies the fraction of dense, cold gas turned into stars per free-fall time is low, typically around a few per cent \citep[see e.g.][]{Krumholz2012} and this simple star formation model easily captures the observed scaling of star formation rate with gas surface density \citep[see e.g.][for a concise overview]{2008MNRAS.383.1210S}.  

The ability to follow low mass stellar units or even individual stars in galaxy evolution simulations has changed the focus of numerical studies to entire galactic star cluster populations. Together with observationally well determined galactic star cluster properties \citep[see e.g.][for reviews]{ZwartReview2010, 2019ARA&A..57..227K} this has opened a new diagnostic window for the small scale structure of the star forming gas and the impact of stellar clustering in galaxy evolution simulations. Star cluster population studies therefore support the scientific validation or falsification of current and novel future theoretical models for the evolution of galaxies with resolved ISM properties. 

The origin and impact of clustered star formation is a fundamental question in star formation studies. 
From observations, we observe star formation to be clustered, although the fraction of stars born in clusters heavily depends on the definition of the cluster \citep{2010MNRAS.409L..54B, 2011MNRAS.410L...6G}. Star clusters are observed wherever there is star formation, irrespective of galaxy mass, such as the Small Magellanic Cloud, the Milky Way, or the Antennae galaxies. The stellar clusters are observed to follow uniform cluster mass functions (CMFs) $\mathrm{dN/dM \propto M^{\alpha}}$ with power law slopes of around $\alpha \sim -2$. The observed normalisations of the CMFs change with the star formation rate of the galaxies and the age of the cluster populations \citep{Fall2012}.  Low mass clusters typically disperse quickly, while young massive clusters (YMCs) that we observe today embedded in the ISM might be longer lived and have properties which could make them potential globular cluster progenitors \citep{Longmore2014,2019ARA&A..57..227K}.

A fundamental observed property of star clusters is that the number of clusters in star forming galaxies decrease with the age of the cluster population but the slope of the mass function is unchanged \citep[see e.g.][]{2019ARA&A..57..227K}. There are discussions in the literature whether the cluster formation rate (CFR) follows the global star formation rate, irrespective of galaxy type, meaning that the fraction of stars born in clusters is independent of the star formation rate per unit area \citep[e.g.][]{Chandar2015,Chandar2017}. Or on the other hand, young star clusters 
show higher rates of disruption in galaxies with higher gas densities and star formation rates \citep{2012MNRAS.419.2606B} but the fraction of stars born in clusters increases for high star formation rates per unit area \citep[e.g.][]{Kruijssen2012,2012MNRAS.419.2606B,Adamo2020}.


In almost all recent high-resolution galaxy evolution simulations, the normalisation of the star formation rate is regulated by stellar feedback. It becomes independent of the assumed \sfe\ for the dense star forming gas \citep[see e.g.][and many other simulations thereafter]{2011MNRAS.417..950H}. However, variations of \sfe\ can change the structure and distribution of newly formed stars, such as cluster sizes, cluster mass functions, and the fraction of stars formed in clusters. 

Recently, some of the highest resolution galactic studies have focused on clustering/star cluster properties in galaxy evolution simulations. \citet{2015MNRAS.446.2038R} report clusters above $10^5$ \solarmass\ with typical sizes of $\sim$ 5 pc without a clear indication for power-law cluster mass functions. \citet{2017ApJ...834...69L} use a cluster formation model combined with a free-fall based star formation model. They find power law like mass functions for clusters above $\sim 10^3$ \solarmass\ and cluster formation efficiencies increasing with star formation rate surface densities. In a follow-up study \citet{2018ApJ...861..107L} found that the global galactic properties are almost insensitive to changes in \sfe\ as long as the feedback is sufficient. For low values of \sfe\ $\leq 0.1$ their cluster age spreads are inconsistently larger than predicted by current observations. They conclude that the range \sfe\ $ = 0.5-1.0$ matches observations best. The cluster formation model, however, does not allow for an investigation of the internal cluster structure.

Assuming a very high local star formation efficiency of \sfe\ $= 1$, \citet{2020MNRAS.493.4315M} report cluster mass functions with slope $\alpha \sim -2$ above $10^{4.5}$ \solarmass\ and typical sizes of $\sim 20$ pc. These sizes are larger than for observed YMCs in the nearby Universe. High local star formation efficiencies are plausible for dense regions of star forming clouds due to the resemblance of the cloud core mass function and the stellar initial mass function \citep[e.g.][]{Wu2010, Evans2014, Heyer2016, Lee2016, Vutisalchavakul2016}. In the simulations, "early" stellar feedback before SN explosions then regulates the global efficiencies down to observed values \citep{2020MNRAS.491.3702H}. However, \citet{2020MNRAS.493.4315M} report more stellar mass in bound clusters i.e. higher cluster formation efficiencies, in simulations with lower star formation efficiencies. In a recent study, \citet{2021MNRAS.506.3882S} conclude that the formation of HII regions has the strongest impact on the clustering of SN explosions and the results are independent of the assumed star formation efficiency parameter. \citet{2021MNRAS.501.5597G} also find that SN feedback reduces the clustering of young stars.

On the other hand, \citet{2018ApJ...861....4S} find a dependence of the galactic depletion time with star formation efficiencies lower than $\sim$ 1 per cent and a decrease of the fraction of star forming gas for efficiencies higher than $\sim$ 1 per cent. 
In a related study, \citet{2021arXiv210313406S} test the effect of variable \sfe\ and fixed \sfe\ on the observed spatial de-correlation between star formation and molecular gas \citep[e.g.][]{2019Natur.569..519K} with the conclusion that low ($\sim$ 1 per cent)/high efficiencies under-/over-predict the spatial decorrelation.  

For some of the highest resolution simulations, \citet{Lahen2019} and \citet{Lahen2020} using dwarf merger simulations with 4 \solarmass\ resolution find clear evidence for power-law star cluster mass functions from a few hundred to $\sim 10^6$ \solarmass\ with increasing formation efficiency in regions with high gas and star formation rate surface densities. At this resolution, also a first investigation of the internal star cluster rotation/kinematics has become possible \citep{2020ApJ...904...71L}. These studies have assumed \sfe\ $= 0.02$ together with a Jeans threshold for immediate star formation. While many star cluster properties for massive clusters ($\gtrsim 10^4$ \solarmass) are in good agreement with the observations, the entire star cluster population appears for be more compact than observed, making the observed cluster disruption difficult. In contrast, the results of the model presented in \citet{2017MNRAS.464.3580D} show clear evidence for rapid cluster disruption. However, their simulated clusters have lower densities than observed clusters at a resolution of $\sim$ 300 \solarmass\ per particle. This would artificially support tidal disruption. 

In summary, no high-resolution simulation so far has produced star clusters with formation properties and an evolution history (i.e. disruption) that are in agreement with observations. While power-law mass functions seem to be a general outcome, the star clusters are either too compact and do not dissolve, or when they dissolve they have been too diffuse at their formation. 

The aim of this study is to investigate the effect of the star formation efficiency per free-fall time on the properties of star cluster populations in dwarf galaxies. Changing this parameter effectively controls how dense a collection of gas particles is allowed to become before star formation begins, along with the associated stellar feedback. We present a suite of isolated gas-rich dwarf galaxy simulations. These simulations have a gas particle mass resolution of 4 \solarmass\ and realise individual massive stars with their respective evolutionary tracks as well as modelling their radiation and supernova feedback at sub-parsec spatial resolution. This allows us to realise individual clusters down to the smallest observed cluster masses of $\sim$ 200 M$_\odot$ in order to examine the effect of varying \sfe\ on the cluster properties and global galaxy properties. 

The simulation suite is a part of the \textsc{Griffin} project\footnote{\url{https://wwwmpa.mpa-garching.mpg.de/~naab/griffin-project}} (Galaxy Realizations Including Feedback From INdividual massive stars). The aim of this project is to perform galaxy scale simulations of individual galaxies and galaxy mergers \citep[e.g.][]{Lahen2020} at such high resolution and physical fidelity that individual massive stars can be realised and important feedback processes such as supernova explosions \citep{Steinwandel20} can be reliably included to study the formation of a realistic non-equilibrium multi-phase interstellar medium \citep{Hu2016, Hu2017, Hu2019}. As discussed in \citet{NaabOstriker2017}, the level of detail included in modern numerical simulations is of significant importance as the environmental density of supernova explosions is controlled by stellar clustering as well as stellar feedback processes.

The paper is organised as follows. Sec. \ref{sec:Methods} describes the simulation setup, particularly the star formation model. We describe the global galaxy properties of the simulations in Sec. \ref{sec:GalaxyProperties} such as the ambient density of star formation and supernovae explosions. We then describe the star cluster analysis. Sec. \ref{sec:StarClusterProperties} describes how we identify friends-of-friends (FoF) groups and perform an energetic unbinding routine in order to identify bound clusters. We then analyse these FoF groups and bound clusters with a discussion of the CMF, cluster formation efficiency (CFE), ages and sizes. We contrast and discuss these findings on both global and small scales in Section \ref{sec:Discussion}, before summarising our findings in Section \ref{sec:Conclusions}.

\section{Methods}
\label{sec:Methods}

\subsection{Simulation code}
\label{sec:SimulationCode}
All simulations were run using a modified version of the smoothed particle hydrodynamics (SPH) code SPHGal presented in \citet{Hu2014,Hu2016,Hu2017}, based on \textsc{Gadget-3} \citep{Springel05}. SPHGal is a well tested implementation developed to more appropriately treat fluid mixing, alleviating many of the previously studied difficulties of SPH codes. Gas dynamics are modelled using a pressure–energy formulation \citep[see e.g.][]{Read10,SaitohMakino13}, with the gas properties smoothed over $N_{\mathrm{ngb}} = 100$ neighbouring particles using the Wendland C$^4$ kernel \citep{Wendland95, Dehnen2012}. A `grad-h' correction term \citep{Hopkins13} ensures better conservation properties in regions of strongly varying smoothing lengths. The artificial viscosity modelling is updated to better account for converging flows and shear flows \citep{Monaghan97,Springel05,CullenDehnen10}. SPHGal also includes  artificial conduction of thermal energy in converging gas flows to suppress internal energy discontinuities. Time-stepping is augmented with a limiter to keep neighbouring particles within a time step difference by a factor of four to capture shocks, in particular from SN explosions accurately \citep[see e.g.][]{SaitohMakino09, DurierDallaVecchia12}. For a more detailed explanation, please see \citet{Hu2014,Hu2016,Hu2017}. 

\subsection{Initial conditions}
\label{sec:InitialConditions}
All simulations presented in this paper are produced from identical initial conditions, described in \citet{Hu2016}. The initial conditions were set up using the method developed in \citet{Springel05}. The dark matter halo follows a Hernquist profile with an NFW-equivalent \citep{NFW97} concentration parameter $c = 10$ with a virial radius $\mathrm{R_{vir} = 44\ kpc}$ and a virial mass $\mathrm{M_{vir} = 2 \times 10^{10}\ M_{\odot}}$. Embedded in this dark matter halo is a $\mathrm{2 \times 10^7\ M_{\odot}}$ stellar disk as well as a $\mathrm{4 \times 10^7\ M_{\odot}}$ gas disk. The initial disk consists of 4 million dark matter particles, 10 million gas particles and 5 million stellar particles, setting a dark matter particle mass resolution of $m_\mathrm{DM} = 6.8 \times 10^3$ \solarmass\ and a baryonic particle mass resolution of $m_\mathrm{baryonic} =$ 4 \solarmass. The gravitational softening lengths are $\epsilon_\mathrm{DM} = 62$ pc and $\epsilon_\mathrm{baryonic} = 0.1$ pc for the dark matter and baryonic particles, respectively.

In this paper, we present seven simulations, all with identical initial conditions. For six of these simulations, we vary their star formation efficiency per free-fall time \sfe\ between 0 and 50 per cent, which we refer to as \zero, \ptwo, \two, \ten, \twenty\ and \fifty. We also ran our fiducial model \two\ without photoionisation, which we refer to as \nophoto. For convenience we refer to the simulations with their star formation efficiency percentages in the figures.

\subsection{Chemistry}
\label{sec:Chemistry}
Our model for chemistry and cooling closely follows the implementation in the \textsc{SILCC}\footnote{\url{https://hera.ph1.uni-koeln.de/~silcc/}} and the \textsc{GRIFFIN} project \citep{Walch15, Girichidis16,Hu2016,Lahen2019}, based on earlier work by \citet{NelsonLanger97, GloverMacLow07, GloverClark12}. We track the chemical composition of gas and stars by following the abundances of 12 elements (H, He, N, C, O, Si, Mg, Fe, S, Ca, Ne and Zn) based on the implementation in \citet{Aumer13}, as well as the non-equilibrium evolution of six chemical species (H$_2$, H$^+$, H, CO, C$^+$, O) and free electrons. The abundances of the first three species are integrated explicitly based on the rate equations within the chemistry network. H$^+$ is formed via collisional ionisation of hydrogen with free electrons and cosmic rays, and is depleted through electron recombination in the gas phase and on the surfaces of dust grains. H$_2$ is formed on the surfaces of dust grains and destroyed via interstellar radiation field photodissociation, cosmic ray ionization and collisional dissociation with H$_2$, H and free electrons.

\subsection{Cooling \& heating}
\label{sec:CoolingandHeating}
We use a set of non-equilibrium cooling and heating processes, where the processes depend on the local density and temperature of the gas as well as the chemical abundance of species, which may not be in chemical equilibrium. Cooling processes include fine structure lines of C$^+$, O and Si$^+$, the rotational and vibrational lines of H$_2$ and CO, the hydrogen Lyman $\alpha$ line, the collisional dissociation of H$_2$, collisional ionization of H, and the recombination of H$^+$. Heating processes include photo-electric heating from an interstellar radiation field, generated by new stars, varying in space and time \citep{Hu2017}, photoelectric effects from dust grains and polycyclic aromatic hydrocarbons, ionization by cosmic rays, photodissociation of H$_2$, ultraviolet (UV) pumping of H$_2$, and the formation of H$_2$. 
For high-temperature regimes, where $\mathrm{T>3\times 10^4 K}$, the simulations do not follow non-equilibrium cooling and heating processes. Instead, we adopt a cooling function presented in \citet{Wiersma09} which assumes an optically thin interstellar medium (ISM) that is in ionization equilibrium with a cosmic UV background from \citet{HaardtMadau01}.

\subsection{Star formation model}
\label{subsec:StarFormationModel}
The star formation algorithm samples stellar masses from a Kroupa IMF \citep[][]{Kroupa01} with an upper limit of 50 M$_\odot$. Sampled masses greater than the gas particle mass (4 M$_\odot$) are statistically realised as individual stellar particles. Should the sampled mass be greater than the gas particle mass, the remaining mass is taken from nearby star forming gas particles in order to conserve mass in the simulation. Sampled masses lower than 4 M$_\odot$ are realised as stellar population particles that store the IMF constituents with a mass above 1 \solarmass\ \citep[see][]{Hu2016}.

Every gas particle has an associated Jeans mass, defined as
\begin{equation}
    M_{J} = \dfrac{\pi^{5/2}c_{s}^3}{6G^{3/2}\rho^{1/2}},
\end{equation}
where $c_s$ is the local sound speed of the gas, $G$ is the gravitational constant and $\rho$ is the gas density.

A gas particle becomes defined as `star-forming' only if $M_{J} < N_{\mathrm{thres}}M_{\mathrm{ker}}$, where $M_\mathrm{ker} = N_\mathrm{ngb}m_\mathrm{gas}$ is the SPH kernel mass and $N_{\mathrm{thres}}$ is a free parameter. As in \citet{Hu2017}, we adopt $N_{\mathrm{thres}} = 8$ to properly resolve the Jeans mass for the star-forming gas.

For gas particles with Jeans masses between 0.5 $M_\mathrm{ker}$ and 8 $M_\mathrm{ker}$ we use a `Schmidt-type' \citep{Schmidt1959} approach to calculate the local star formation rate:
\begin{equation}
    \mathrm{\dfrac{d\rho_*}{dt}} = \sfe \dfrac{\rho_\mathrm{gas}}{t_\mathrm{ff}},
    \label{eq:localSFR}
\end{equation}
where $\mathrm{\epsilon_{ff}}$ is the star formation efficiency per free-fall time, $t_\mathrm{ff} = \sqrt{3\,\pi/(32\, G\,\rho_\mathrm{gas})}$  is the gas free-fall time \citep{BinneyTremaine2008}, and $\rho_*$ and $\rho_\mathrm{{gas}}$ are the stellar and gas volume densities respectively. For gas particles with a $M_J < 0.5$, we enforce instantaneous star formation, as introduced in \citet{Lahen2019}.  

In this study we explore the effect of varying the star formation efficiency parameter \sfe\ on the formation of star clusters.
In equation (\ref{eq:localSFR}), \sfe\ is the fraction of `star-forming' gas which is turned into stars after a gravitational free-fall time. A unit efficiency \sfe\ = 1 describes a star formation rate for which all local star-forming gas is converted into stars on a free fall time. This numerical implementation has its origin in the early days of numerical galaxy formation simulations \citep[see e.g.][]{Katz92} and was motivated by galaxy observations \citep[e.g.][]{Kennicutt1998} on kpc scales. Despite higher resolution and physical fidelity of the simulations this star formation model is still being used \citep[see e.g.][]{Semenov21}. This can be motivated by the finding that a relation between star formation rate and gas density is also valid within star-forming clouds \citep[see e.g.][]{Pokhrel21} 

By varying  \sfe, we control how much a region of self-gravitating gas particles is allowed to collapse before stars begin to form. 
For a collapsing cloud, it is expected that high values of \sfe\ allow gas particles to be converted into stars while the cloud is still relatively diffuse. Stellar feedback in the form of radiation and SN explosions is more efficient at low densities and might easily disperse the clouds before reaching high densities. 
A low value of \sfe\ allows the gas to collapse to higher densities before forming stars. This will generate denser stellar systems and stellar feedback might be less efficient at gas dispersal.

\subsection{Stellar Feedback}
\label{subsec:StellarFeedback}
In lower resolution galaxy formation simulations, a star particle typically represents an entire population of stars with an assumed IMF \citep[see e.g.][for a review]{NaabOstriker2017}. From this, the abundance of massive stars is calculated and subsequently the energy budget of the stellar feedback of each stellar population particle. For the Kroupa IMF used in this study \citep{Kroupa2001}, there is around one type II supernova per 100 \solarmass\ of formed stars. A given stellar population particle with mass, m$_*$ would inject (m$_*$/100 \solarmass) $\times 10^{51}$ erg into the ISM. In the simulations presented here however, we assume a minimum stellar mass of 4 \solarmass\ representing the low mass part of the IMF. Every massive star expected to form from the assumed IMF is realised individually in the simulation. Therefore, all stars that explode as SN are realised individually. As mentioned in Sec. \ref{subsec:StarFormationModel}, particles below 4 \solarmass\ are realised as stellar population particles with IMF constituents drawn from an IMF with a mass above 1 \solarmass.

At this mass resolution and 0.1 pc gravitational force softening in our simulations, individual SNe are well resolved at ambient densities $\mathrm{n < 10\ cm^{-3}}$ \citep[see e.g.][Appendix B]{Hu2016}. As we show in Section \ref{subsec:ambientdensitySNII} this corresponds to more than 99 per cent of all SNe in our simulations with photoionisation, and more than 97 per cent for \nophoto. Explosions at higher ambient density do not capture all details of the Sedov phase but result in the input of the expected amount of radial momentum to the ambient ISM \citep[see][]{Hu2016,Steinwandel20}.

We model the photoionisation of hydrogen (PI) by massive stars with a Str{\"o}mgren approximation assuming the recombination rates balancing the photon production rates. The PI model reproduces well the evolution of D-type fronts \citep{Spitzer1998} in good agreement with the \textsc{Starbench} results for different numerical implementations \citep{Bisbas2015}. We note that this model is a good approximation for the local impact of hydrogen ionising radiation but does not accurately follow the radiation field in dwarf galaxies on larger scales \citep[see][for a discussion]{Emerick2018}. 

\begin{figure*}
\includegraphics[width=0.90\textwidth]{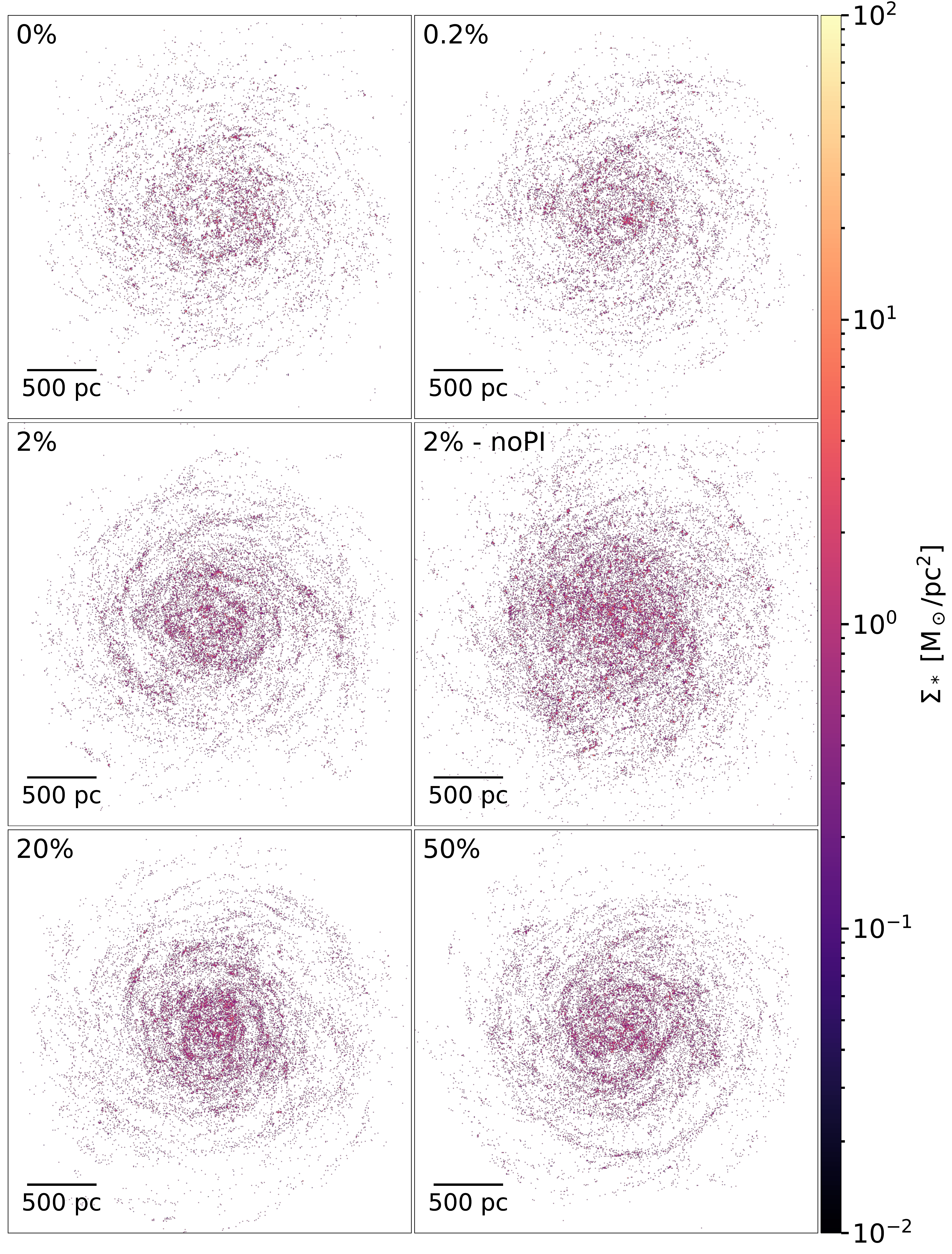}
\caption{Face-on distribution of newly formed stars after 400 Myr for six simulations with increasing star formation efficiency from top left to bottom right. For simulation \two\ (middle left) we also show the version without the photo-ionisation model, \nophoto, (middle right). The stellar surface densities are color coded by \solarmass/pc$^{2}$. Each image shown is 3$\times$3 kpc$^2$ plotted with 1024$\times$1024 pixels. Dense and compact star clusters form in the low efficiency simulations with half-mass surface densities as high as 5 $\times$ 10$^3$ \solarmass/pc$^{2}$. The most extreme case is \nophoto, where we see surface densities as high as 10$^4$ \solarmass/pc$^{2}$. At high star formation efficiencies (\twenty\ and \fifty, bottom row) the stellar distribution is significantly smoother with fewer and more diffuse visible clusters. This showcases that star formation model parameters and feedback models have a strong impact on stellar clustering.}
\label{fig:ALLeffs_6panels_stars_snap400}
\end{figure*}


\begin{figure*}
\includegraphics[width=0.90\textwidth]{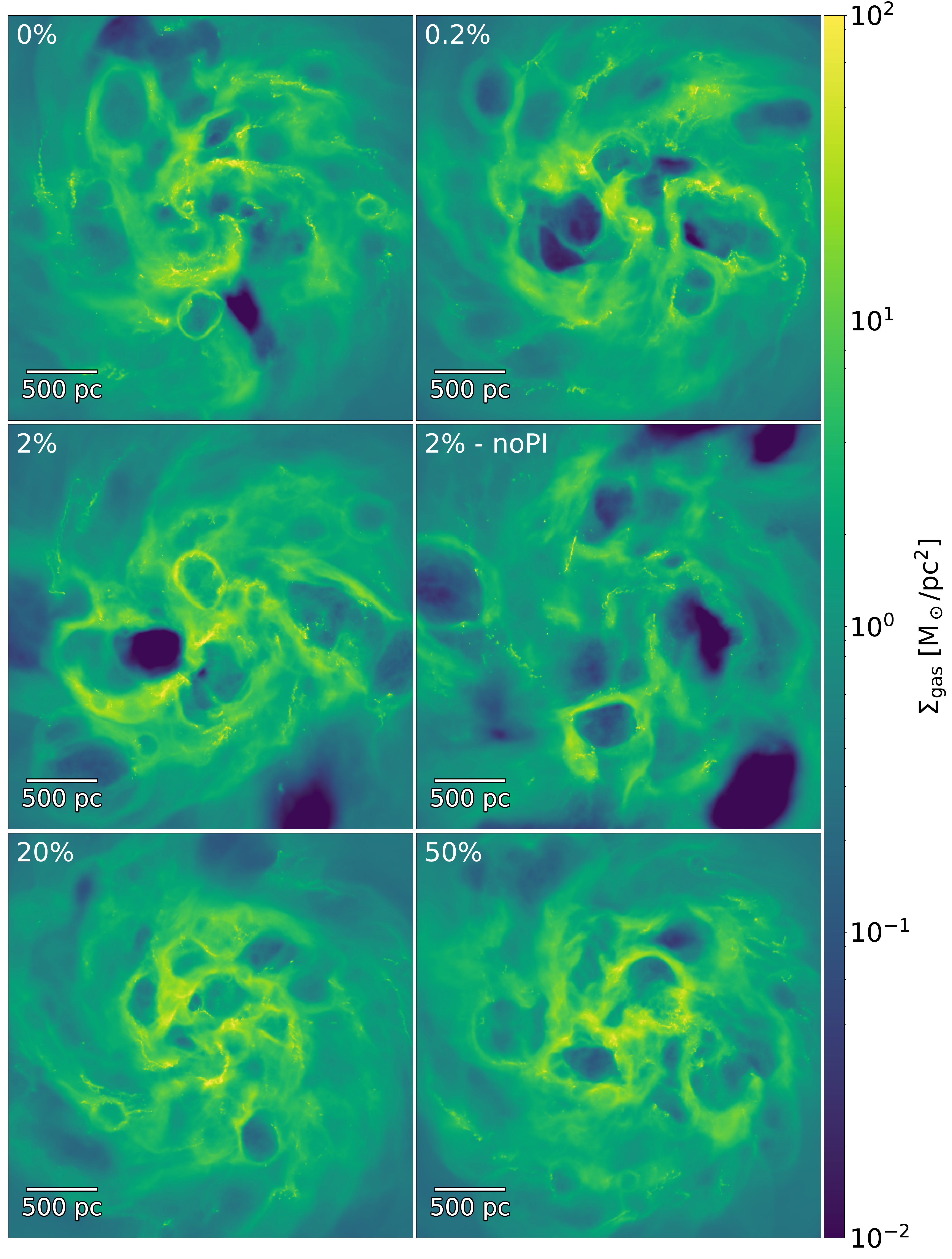}
\caption{Face-on gas surface densities of the six simulations at time t = 400 Myr  (see Fig. \ref{fig:ALLeffs_6panels_stars_snap400} for the stellar distribution). The gas is structured in a diffuse component, dense filaments, and shells generated by photo-ionisation and SN explosions. All simulations with photo-ionisation show similar structure. The \nophoto\ model (middle right panel) has less dense gas which is dispersed by the strongly clustered SN feedback from the forming star clusters.}
\label{fig:ALLeffs_6panels_gas_snap400}
\end{figure*}


\section{Galaxy properties}
\label{sec:GalaxyProperties}


\begin{figure}
\includegraphics[width=\columnwidth]{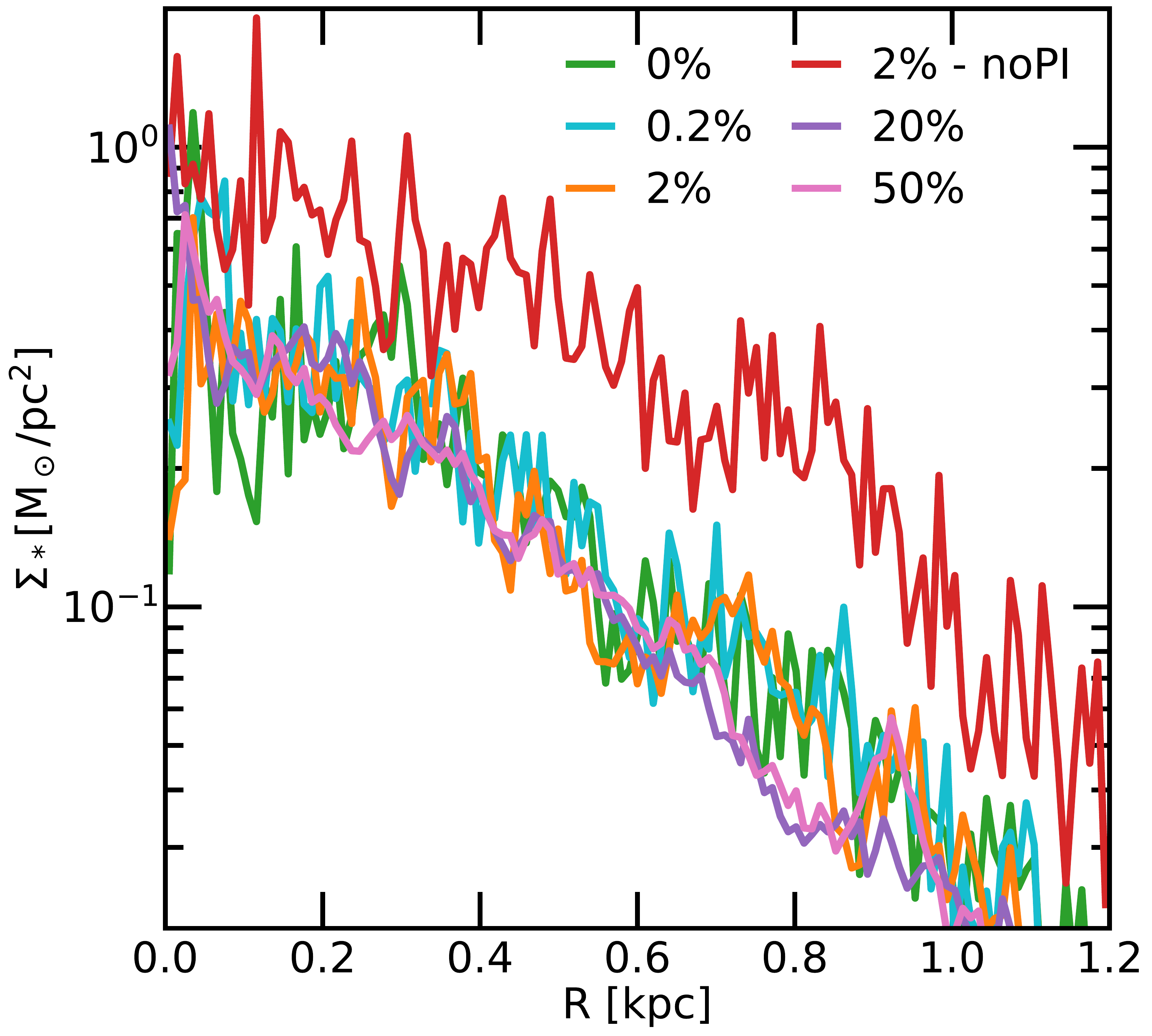}
\caption{Radial stellar surface density profiles for new stars at 400 Myr for each simulation out to 1.2 kpc. The profiles are smoothed over a spatial scale of 10 pc. All simulations apart from \nophoto\ have a similar radial structure. The higher star formation efficiencies result in smoother radial profiles which can also be seen qualitatively in Fig. \ref{fig:ALLeffs_6panels_stars_snap400}.}
\label{fig:RadialProfiles_singleplot_ALLeffs_ext1200_snap400}
\end{figure}

\subsection{Global properties}
\label{subsec:GlobalProperties}
In Figs. \ref{fig:ALLeffs_6panels_stars_snap400} and \ref{fig:ALLeffs_6panels_gas_snap400} we show the face-on distribution of stars and gas after 400 Myr simulation time for our six simulations. Visually, Fig. \ref{fig:ALLeffs_6panels_stars_snap400} highlights the differences in the distributions of stars as we increase the star formation efficiency (from the top left panel down to the bottom right panel) as well as the effect of removing the photo-ionisation model (middle right panel). We can see that for low star formation efficiencies (top row), the newly formed stars are more clustered. In contrast, the stellar distributions for the higher star formation efficiency simulations (bottom row) are more smooth, with less clustered star formation. Despite the differences in the stellar distributions, the corresponding gas distributions shown in Fig. \ref{fig:ALLeffs_6panels_gas_snap400} do not show substantial differences in structure. There might be a slight trend for more low density bubbles in the lower star formation efficiency simulations (top row) in comparison to the higher star formation simulations (bottom row). We also show the 2 per cent model but without photo-ionisation, \nophoto\ in the middle right panel of both Figs. \ref{fig:ALLeffs_6panels_stars_snap400} and \ref{fig:ALLeffs_6panels_gas_snap400}. Here we see significantly stronger clustering in the stellar distribution compared to the corresponding simulation including photo-ionisation, \two. In addition, about a factor of three more mass in stars has formed in the \nophoto\ simulation by t = 400 Myr: $\mathrm{M_{*}(\nophoto) = 1.22 \times 10^6\ M_\odot}$ vs. $\mathrm{M_{*}(\two) = 3.87 \times 10^5\ M_\odot}$. More gas mass has been used up by star formation in the \nophoto\ run, which is also reflected in the gas distribution in Fig. \ref{fig:ALLeffs_6panels_gas_snap400} where we see lower densities in the gas overall as well as more substantial low density bubbles created by strongly clustered SN feedback.

In Fig. \ref{fig:RadialProfiles_singleplot_ALLeffs_ext1200_snap400} we show the stellar surface density radial profile of all six simulations at 400 Myr, smoothed over 10 pc. The stellar distributions for the runs with photo-ionisation are very similar, but comparing for example the lowest star formation efficiency \zero\ with the highest star formation efficiency \fifty, we can see that the stellar distribution for the \fifty\ simulation is much smoother. For the same physical model, varying the star formation efficiency therefore does not alter the radial distribution of stars. Instead, it alters how smooth the stellar distribution is. For the \nophoto\ run however, we quantitatively see the more efficient transformation of gas into stars by an increased normalisation of the profile. We can also observe the strong clustering in the fluctuations of the surface density.

\begin{table*}
 \centering
 \begin{tabular}{lcccccccccccc} 
  \hline
   Name & \sfe & $\mathrm{{SFR}}$ & $\eta$ & $f_\mathrm{SN,A}$ & $f_\mathrm{SN,B}$ & $\alpha$ & $\Gamma_\mathrm{FoF}$ & $\Gamma_\mathrm{BC}$ & Age Spread & $\mathrm{r_{1/2}}$ & $\mathrm{\overline{M_{5mm}}}$ & f$\mathrm{_{bound}}$ \\ 
    &  & $[10^{-4}$ \solarmass/yr] & & & & &  & & [Myr] & offset [pc] & [$\mathrm{M_\odot}$] & \\
  \hline
  \zero & \textbf{0 $\%$}  & 10 / 8.4  & 51 / 57  & 0.64 & 0.36  & -2.9 $\pm$ 0.3 & 0.61 & 0.57  & 2.5 & -0.7 & 1933 & 0.98\\
  
  \ptwo & \textbf{0.2 $\%$} & 10 / 8.6 & 47 / 52 & 0.64 & 0.36 & -2.9 $\pm$ 0.3  & 0.59 & 0.54 & 2.5 & -0.7 & 2813 & 0.99\\
  
  \two & \textbf{2 $\%$} & 9.2 / 8.6  & 54 / 59 & 0.69 & 0.31  & -2.8 $\pm$ 0.3 & 0.34 & 0.26 & 2.6 & -0.4 & 1631 & 0.88\\
  
  \nophoto & \textbf{2 $\%$ - noPI} & 25 / 5.6  & 32 / 124 & 0.48 & 0.52 & -3.1 $\pm$ 0.3 & 0.64 & 0.62 & 3.7 & -1.1 & 5562 & 0.97\\
  
  \ten & \textbf{10 $\%$} & 7.9 / 7.6 & 50 / 55 & 0.74 & 0.26 & -1.8 $\pm$ 0.3 & 0.10 & 0.02 & 8.1 & 3.5 & 526 & 0.34\\
  
  \twenty & \textbf{20 $\%$} & 7.9 / 7.2 & 49 / 52 & 0.75 & 0.25 & -2.1 $\pm$ 0.2 & 0.10 & 0.01 & 6.2 & 4.9 & 280 & 0.30\\
  
  \fifty & \textbf{50 $\%$} & 7.9 / 7.5 & 53 / 56
  & 0.75 & 0.25  & -1.7 $\pm$ 0.2 & 0.11 & 0.02 & 6.6 & 8.0 & 291 & 0.41\\
  
  \hline
 \end{tabular}
 \caption{Summary of simulation properties. \sfe: star formation efficiency per free-fall time, no-PI does not allow for HII regions; SFR: average star formation rate calculated between 0-500 Myr / 200-500 Myr; $\eta$ average mass loading ($\mathrm{\overline{OFR}/\overline{SFR}}$) calculated between 0-500 Myr / 200-500 Myr; $f_\mathrm{SN,A/B}$: fraction of SN exploding at densities higher (A) and lower (B) than $\mathrm{\rho_{ambient} = 10^{-2}\ cm^{-3}}$; $\alpha$ is the slope of the power-law cluster mass function; $\Gamma_\mathrm{FoF}$ and $\Gamma_\mathrm{BC}$:  average cluster formation efficiencies for FoF groups and bound clusters (BC) with ages < 10 Myr; Age Spread: average age dispersion of bound clusters; $\mathrm{r_{1/2}}$ offset: offset of the half-mass radius from the \citet{2021arXiv210612420B} relation shown in Fig. \ref{fig:MassDensity3Plots_BoundClusters_ALLEFFS_snaps200500}; $\mathrm{\overline{M_{5mm}}}$: average mass of the five most massive clusters; $f_\mathrm{bound}$: fraction of FoF groups identified as bound. \label{table:GalaxyProperties}}
 \end{table*}

\begin{figure}
\includegraphics[width=\columnwidth]{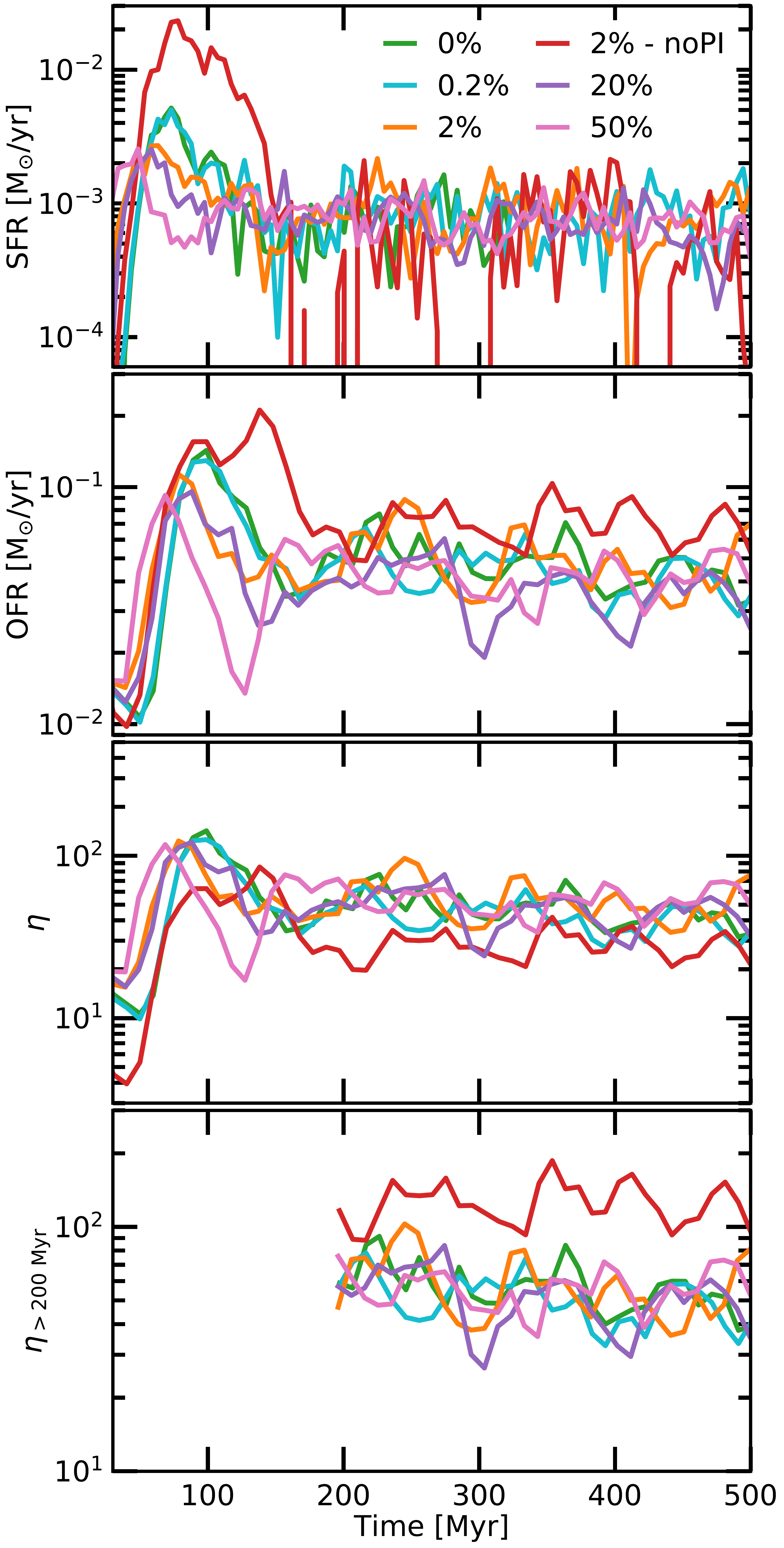}
\caption{\textit{Top panel:} Star formation rates (SFR) for the five different simulations with varying star formation efficiency parameters and the no-photoionisation run. All simulations with photoionisation have similar star formation rates, the one without has a higher average rate and is more bursty. \textit{Second panel:} Gas outflow rate (OFR), which is defined by the gas crossing 500 pc above and below the central disk in 10 Myr intervals. This plot shows relatively similar outflow rates for all simulations at all times during the simulations. In addition we give the average metallicity of the outflowing gas. \textit{Third panel:} Mass loading, $\eta$ defined as the ratio of the outflow rate to the star formation rate. Here we divide the instantaneous OFR by the average SFR across the entire simulation (0-500 Myr). The simulations show no major differences. \textit{Bottom panel:} Mass loading $\eta_\mathrm{>200\ Myr}$ between 200 and 500 Myr. As in the third panel, we show the instantaneous OFR but this time divided by the average SFR between 200-500 Myr, which excludes the initial starburst in each of the simulations. Here we see the mass loading of the run without photo-ionisation (\nophoto) is a factor of two higher than the corresponding run with photo-ionisation (\two).}
\label{fig:SFH_MassLoading_Outflows_ALLeffs}
\end{figure}

Fig. \ref{fig:SFH_MassLoading_Outflows_ALLeffs} shows the star formation rate (SFR, top panel), outflow rate (OFR, second panel) and the mass loading $\mathrm{\eta}$ for all times in the simulation (third panel) as well as for just 200-500 Myr (bottom panel), defined as the ratio of the outflow rate to the average SFR. All simulations including photo-ionisation settle to similar star formation rates. In the first 50-100 Myr, the peak of the onset of star formation is noticeably higher and slightly delayed at lower star formation efficiencies, as it takes time for the gas to reach the density threshold, but as soon as gas manages to reach the density of 0.5 \jeansmass\ (discussed in Sec. \ref{subsec:StarFormationModel}), we immediately form many stars. This high peak in star formation is then reflected in a strong peak in the outflow rate with a small time lag. This feature is seen in all models, but less so in the higher star formation efficiency runs. Star formation is also more bursty in the \zero, \ptwo\ and \two\ runs in comparison to the \twenty\ and \fifty\ runs. The reason for this can be explained by comparing the two extreme models, \zero\ and \fifty. For the \fifty\ run, once the gas passes the upper star formation threshold of 8 \jeansmass, these gas particles are defined as star-forming, and statistically 50 per cent per free-fall time of these star forming gas particles will become stars. In this regime, gas does not have to collapse to high densities before forming stars. Neighbouring gas particles will be affected early in the collapse phase by stellar feedback and supernovae after the formation of a star, resulting in a smoother star formation rate. In contrast, the \zero\ run has no star formation above 0.5 \jeansmass\, and so gas must collapse to much higher densities before forming stars on a short timescale. The feedback from the formed stars will therefore be more effective in keeping neighbouring star forming gas particles away from that density threshold, drastically halting star formation. This subsequently results in more bursty star formation. Despite these fluctuations, the star formation rates for all models with photo-ionisation stay almost constant across the entire simulation, around $10^{-3}$ \solarmass/yr. The reasons for this are discussed next in Sec. \ref{subsec:ambientdensitySNII}.

Along with the SFR, the outflow rates of each of the simulations with photo-ionisation are very similar, remaining around $4-5 \times 10^{-2}$ \solarmass/yr, showing that the substantial differences in the stellar clustering from the different star formation efficiencies does not seem to effect the outflow rates. Subsequently, the mass loading $\eta$ of the simulations with photo-ionisation maintain very similar values, keeping a relatively constant value of approximately $\mathrm{\sim 50-60}$ from 200 Myr onwards. 

For the run without photo-ionisation \nophoto, things look slightly different. Many stars are formed in the first 150 Myr and then from 200 Myr onwards the star formation is much more bursty, with several periods of no/very low star formation. The outflow rate however remains relatively constant at approximately $ 7 \times 10^{-2}$ \solarmass/yr as seen in the second panel. When looking at the average mass loading as calculated across the entire simulation as shown in the third panel, we see a slightly lower value in comparison to the runs including photo-ionisation. However, if we only observe the mass loading after 200 Myr and so excluding the initial starburst, we find the mass loading is approximately a factor of two higher in comparison with the same run but including photoionisation (\two). The values for the SFR and mass loading $\eta$ are summarised in Table \ref{table:GalaxyProperties} where we show both the average values at all times as well as from 200 Myr, excluding any initial starburst in the galaxy. We also check the metallicity of these outflows from each of the simulations and find very little difference. At a typical metallicity of $ Z \sim 0.13\ Z_\odot$, the outflows are metal enriched compared to the initial gas phase metallicity of $Z = 0.1\ Z_\odot$. 


\subsection{The ISM densities for star formation and supernova explosions}
\label{subsec:ambientdensitySNII}

\begin{figure}
\includegraphics[width=\columnwidth]{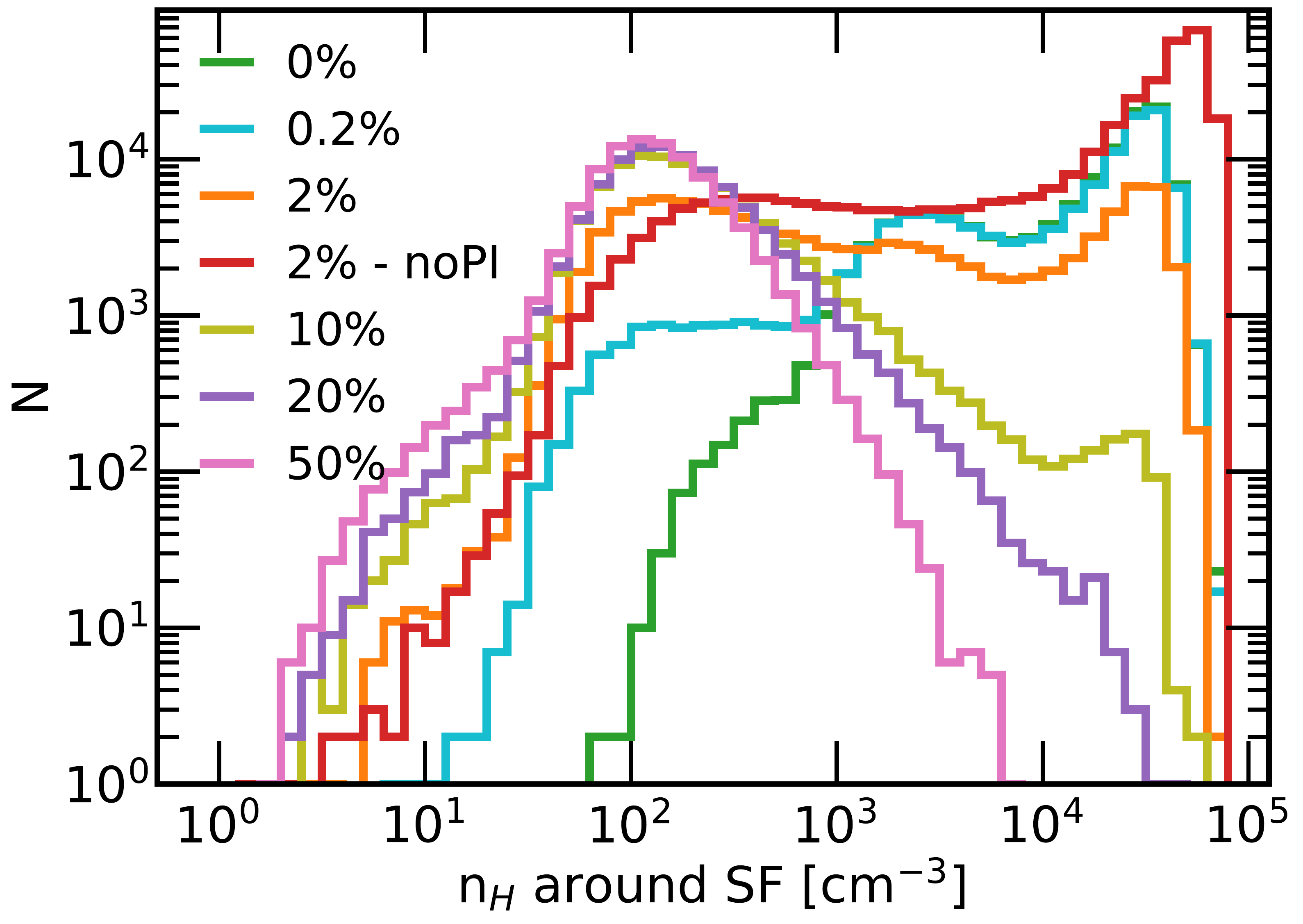}
\caption{Density distribution of gas particles which are turned into stellar particles for all stars present at 400 Myr for all seven simulations. The Jeans threshold of 0.5 \jeansmass\ is traced by the \zero\ simulation (0$\%$, green line).  For \zero, \ptwo, \two\ and \nophoto\ most of the star formation takes place at densities $\mathrm{n_H \gtrsim 10^4\,cm^{-3}}$ with more and more extended tails towards lower densities of $\mathrm{\sim 10^{2}\,cm^{-3}}$. In the high efficiency simulations \ten, \twenty\ and \fifty, most stars form from gas at densities of $\mathrm{n_H \sim 10^{2}\,cm^{-3}}$ and the gas does not even reach the several orders of magnitudes higher threshold densities of e.g. \zero.}
\label{fig:SFdensities_singleplot}
\end{figure}

\begin{figure}
\includegraphics[width=\columnwidth]{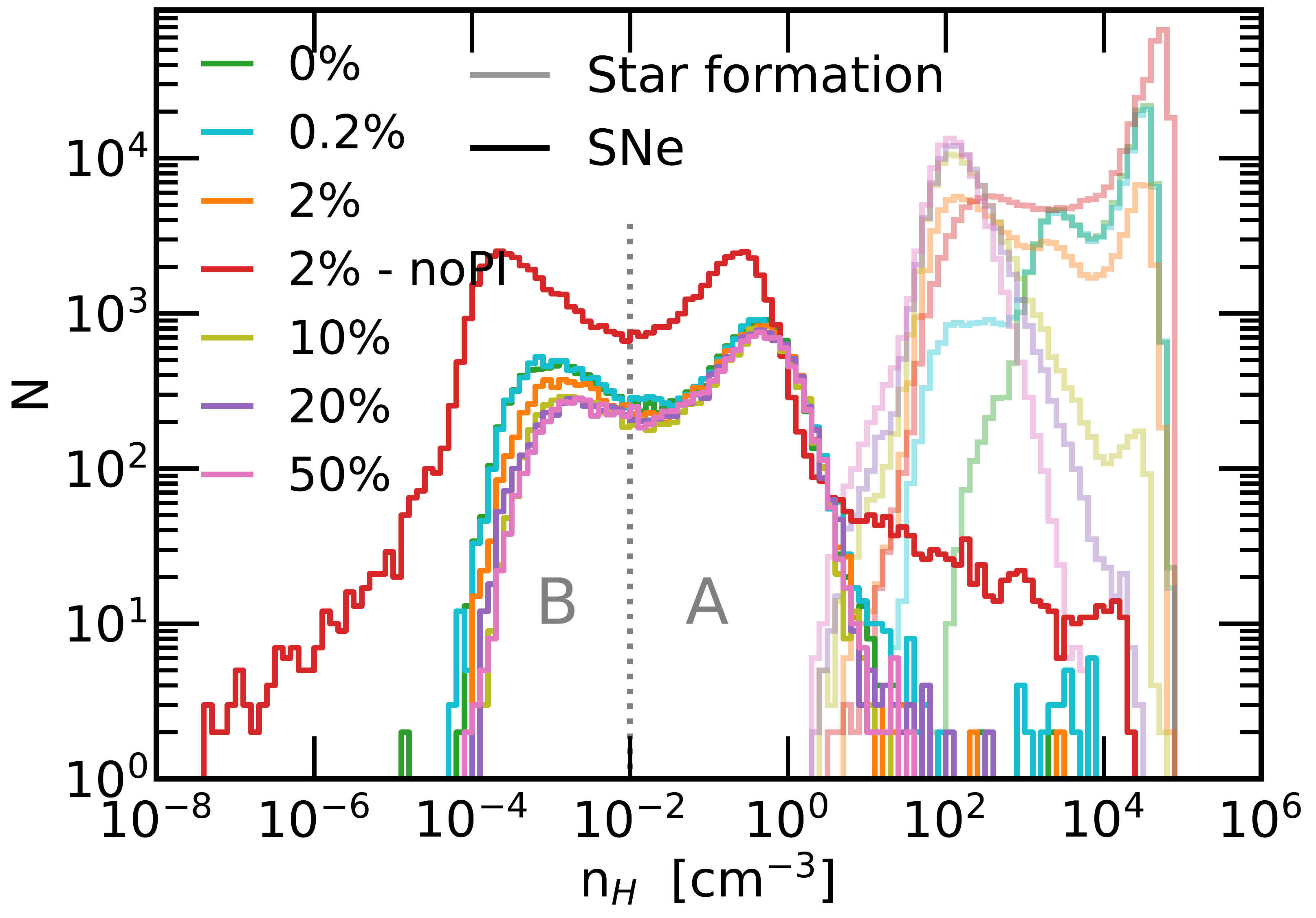}
\caption{Ambient density distributions of SNe explosions within the first 400 Myr (bold lines) compared to the star formation densities from Fig. \ref{fig:SFdensities_singleplot}. The various simulations are indicated by different colors. The SNe explode at several orders of magnitude lower ambient densities than the stellar birth sites with a double peaked distribution, characteristic for all simulations. One peak (A) is at $\sim$ 10$^{-0.3}$ cm$^{-3}$, the second peak (B) is at lower densities of  $\sim$ 10$^{-3}$ cm$^{-3}$, and there is a minimum at  $\sim$ 10$^{-2}$ cm$^{-3}$. The \nophoto\, (2\% - noPI) has a higher SN rate. Less than 2 per cent of the SNe explode at densities higher than $\mathrm{n_H \sim 10\ cm^{-3}}$ indicating that they have no memory of their birth places.}
\label{fig:AmbientDensitySNII}
\end{figure}


\begin{figure}
\includegraphics[width=\columnwidth]{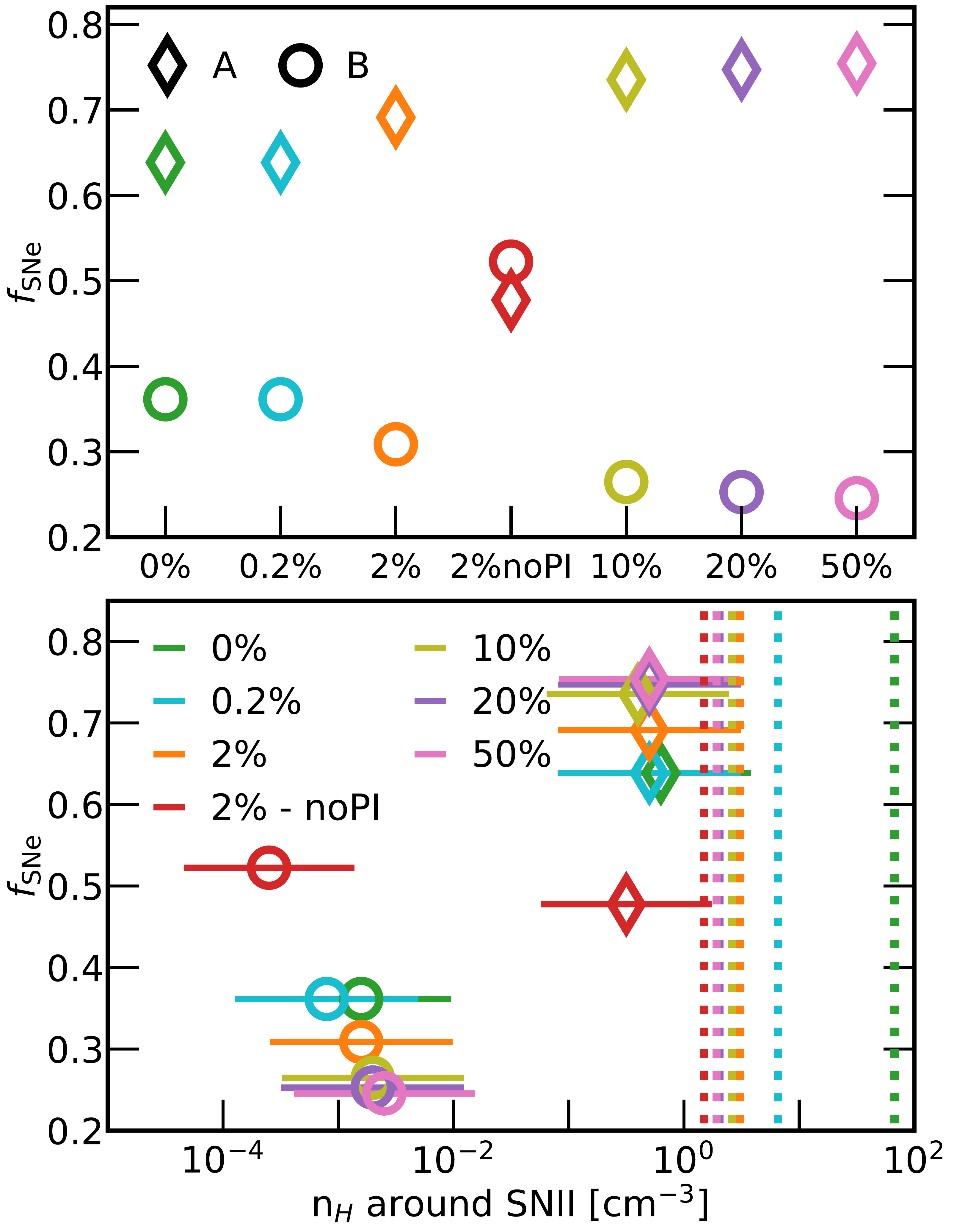}
\caption{\textit{Top panel:} Fraction of SNe exploding at environmental densities higher (A, diamonds) and lower (B, circles) than $10^{-2}$ cm$^{-3}$ (regions A and B in Fig. \ref{fig:AmbientDensitySNII}) for the different simulations. SNe at higher ambient densities dominate. With increasing star formation efficiency the fraction of SNe at low environmental densities decreases. This can be connected to lower cluster formation efficiencies (see Sec. \ref{sec:StarClusterProperties}) and therefore less clustered SN events. The simulation without photo-ionisation (red symbols) show an inverted behaviour with the lower ambient densities becoming dominant. \textit{Bottom panel:} Fraction of core collapse SNe in the high (A, diamonds) and low (B, circles) density regime as a function of ambient density. The colour coding is the same as in the top panel. The symbols indicate the density peaks and the horizontal bars show the dispersion. The dotted vertical lines indicate the lowest star formation densities in each of the simulations (see Fig. \ref{fig:AmbientDensitySNII}). All simulations with photo-ionisation have similar higher (A) and lower (B) density peaks of $\sim$ 10$^{-0.3}$ cm$^{-3}$ and $\sim$ 10$^{-3}$ cm$^{-3}$, respectively. For the  \nophoto\, (2\% - noPI) simulation, the peaks are shifted to slightly lower ambient densities. In contrast to the other simulations, more than half of the SNe explode at lower densities.}
\label{fig:2panel_fits_AmbientDensitySNII}
\end{figure}

Do we have an explanation for why the star formation model presented here results in similar star formation and outflows rates (see Fig. \ref{fig:SFH_MassLoading_Outflows_ALLeffs}) despite the large differences in star formation efficiency? In Fig. \ref{fig:SFdensities_singleplot} we show the density distribution of the gas particles which are transformed into stars. These density distributions vary significantly for different models. The density distribution without a free-fall time based star formation \zero\ (0 \%, green), traces the threshold of 0.5 Jeans masses in the SPH kernel for cold gas from a density of $\mathrm{n_H[T=10 \mathrm{\ K]\approx 5 \times 10^2\,cm^{-3}}}$ to $\mathrm{n_H[T=100 \mathrm{\ K]\approx 5 \times 10^5\,cm^{-3}}}$. The \ptwo\ simulation (0.2 \%, blue) has a similar distribution with an emerging lower density tail due to the additional possibility for lower density gas to experience free-fall time based star formation. Still the gas densities peak in the range $\sim 10^{4}-10^{5} \ \rm cm^{-3}$. In the fiducial \two\ run (2\%, orange) even more star formation at low densities is possible and the distribution becomes broader with a second lower density peak emerging at $\sim 100$ cm$^{-3}$. The corresponding simulation without photo-ionisation (red) shows a similar distribution but with a larger fraction of stars forming at densities higher than $\sim 100$ cm$^{-3}$. The high star formation efficiency runs \twenty\ and \fifty\ (20\%, purple; 50\% pink) do not reach high enough star formation densities to hit the Jeans threshold but form all their stars in the free-fall regime between 8 and 0.5 Jeans masses. As a consequence, star formation mostly takes place at gas densities of $\sim 100$ cm$^{-3}$. Comparing the models as we decrease \sfe\ for example from \fifty\ to \ten, we observe an increase in star formation at higher densities above $10^4$ cm$^{-3}$ and a decrease in the number of stars formed at lower densities, as we would expect. In summary, the ISM density distributions at which the stars form are qualitatively different when the star formation efficiency is varied.

In Fig. \ref{fig:AmbientDensitySNII} we compare the star formation densities shown in Fig. \ref{fig:SFdensities_singleplot}, now repeated as lightly shaded lines, to the distribution of the ambient ISM densities at which the massive stars explode as supernovae shown as the solid lines. In contrast to the star formation densities, all simulations including photo-ionisation show a similar behaviour for the ambient SN densities. The vast majority of the SNe explode at densities lower than the densities of star formation with two peaks: a first peak (A) at $\mathrm{n_H \sim 10^{-0.3} cm^{-3}}$ and a second peak (B) at lower densities of $\mathrm{n_H \sim 10^{-3} cm^{-3}}$. For simplicity we have separated the ambient density distributions at a single fiducial density of  $\mathrm{n_H \sim 10^{-2}}$, which approximately corresponds to the local minimum, into a "high" density region A and a "low" density region B. For the simulations with photo-ionisation, the majority of SNe explode at higher ambient densities (region A) while the number becomes about equal for \nophoto\ (2\% - noPI). The distribution of densities also becomes broader with the exclusion of photo-ionisation. Very few ($\lesssim 2$ per cent) SNe explode at high densities $\mathrm{n_H \gtrsim 10 cm^{-3}}$ while typical stellar birth densities are much higher. An observational study by \citet{Hewitt09} has used masers as signatures of supernovae remnants (SNRs) interacting with molecular clouds. Assuming the survey to be complete, they find around 15 per cent of SNRs are maser emitting. In the simulations, however, we only track the local ambient density at the time of the explosion. Some expanding supernova remnant shells may interact with dense gas thereafter.

In the top panel of Fig. \ref{fig:2panel_fits_AmbientDensitySNII} we show the fraction of SNE exploding in the high and low ambient density regimes A and B for all simulations. More than 60 per cent explode at higher densities and the fraction is increasing for simulations with high star formation efficiencies. The \nophoto\, (2\% - nophoto) simulation shows about equal numbers of SNe in both regions. The bottom panel of Fig. \ref{fig:2panel_fits_AmbientDensitySNII} show the SN fractions as a function of the peak densities in the two regimes. The peak densities are very similar for the simulations including photo-ionisation and slightly lower for the one without. The bars indicate the dispersion in density. The lowest star formation densities indicated by the vertical dashed lines hardly overlap with the densities at explosion time. This indicates that the stars have "forgotten" about their birth environment as soon as they explode as SNe, i.e. the typical massive star explodes in a completely different environment than where it was born and this environment appears to be "universal" and independent of the details of the star formation model. This is a plausible explanation for why the outflow rates of the models with different star formation efficiency are so similar (see Fig.\ref{fig:SFH_MassLoading_Outflows_ALLeffs}) in all models. The SNe couple to the ISM in a very similar way. 

We suggest that most SNe explode at typical ISM densities (region A) for these dwarf galaxy systems, which is dominated by neutral gas in equilibrium. At lower ambient densities (region B in Fig. \ref{fig:AmbientDensitySNII}) SNe explode in pre-processed environments, mostly affected by previous nearby SNe explosions.  Qualitatively, these results agree with previous simulations investigating SN ambient density distributions \citep[e.g.][]{Hu2017,Peters2017}.

\section{Identifying Star Clusters and their Properties}
\label{sec:StarClusterProperties}

In this section we discuss how we identify clusters of stars in our simulations as well as the properties of these clusters in the different simulations. Note here that we primarily only discuss four of the seven simulations previously shown, \ptwo, \two, \twenty\ and \fifty. The cluster properties from the \zero\ simulation are very similar to the \ptwo\ simulation, whilst the \ten\ population is very similar to the \twenty.

We identify clustered stars in the simulations using a friends-of-friends algorithm \citep[FoF, see e.g.][]{Davis85} with a linking length of 5 pc. For each FoF group, which represents a star cluster, we impose a minimum of 35 stars \citep[see e.g.][]{LadaLada2003} as well as a minimum mass of 200 M$_\odot$ for the analysis. Following the FoF analysis, we perform a binding energy analysis as described in Section \ref{subsec:unbindingprocedure} on each of these FoF groups to determine if it is a bound cluster. Throughout this paper, effort is made to distinguish between:
\begin{itemize}
    \item \textit{\textbf{FoF groups}}: Physical associations of stars identified by the friends-of-friends algorithm. Observationally, this corresponds to stellar associations and star clusters.
    \item \textit{\textbf{Bound clusters}}: Bound groups of stars verified by the unbinding procedure described in Section \ref{subsec:unbindingprocedure}. These are observed as bound star clusters.
\end{itemize}

\subsection{Virial Analysis}
\label{subsec:VirialAnalysis}

\begin{figure}
\includegraphics[width=0.95\columnwidth]{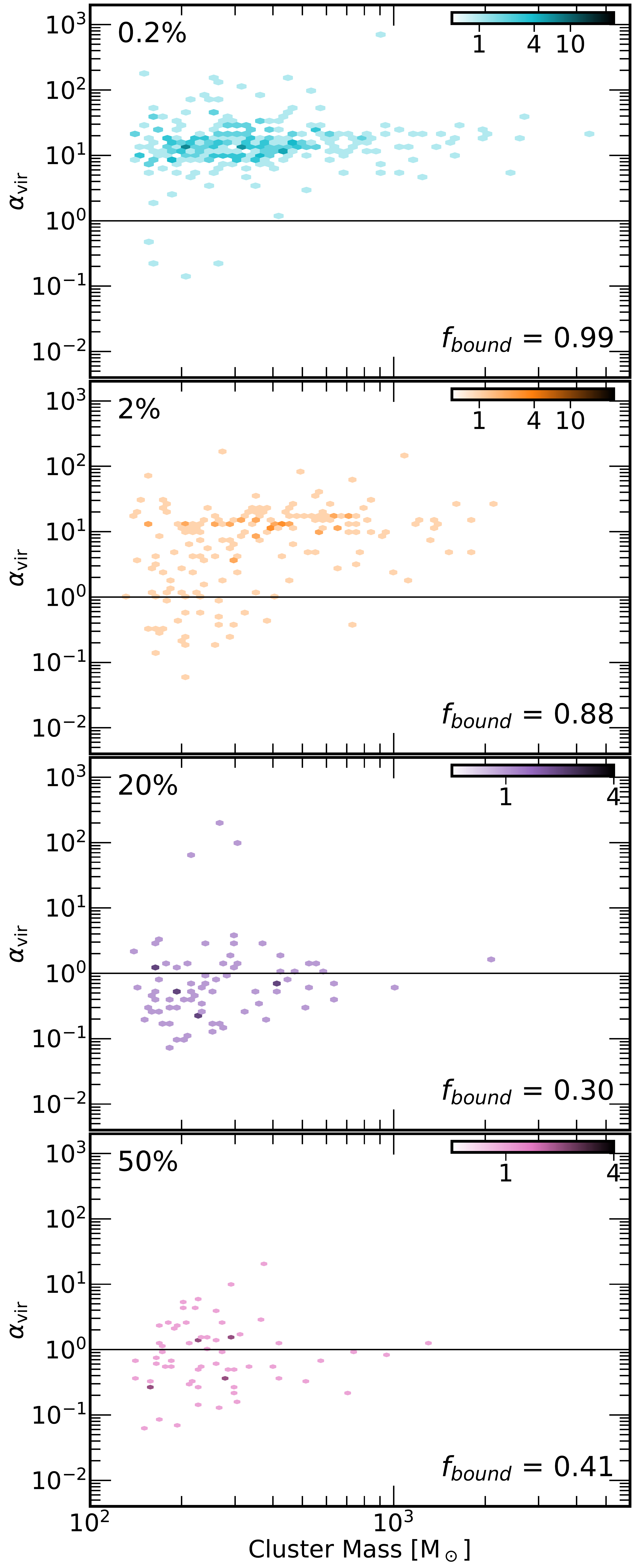}
\caption{Virial parameter $\alpha_{\mathrm{vir}}$ as a function of cluster mass for identified FoF groups in four simulations with increasing star formation efficiency from top to bottom. The colourbar shows the number of clusters in each hexbin. We show all FoF groups with stellar ages younger than 20 Myr between 200 and 500 Myr in 20 Myr intervals. The fraction of bound groups (virial parameter larger than unity, horizontal line), $f_\mathrm{bound}$ decreases from close to unity to $\sim 0.4$ With increasing star formation efficiency. Also the total number of identified stellar groups decreases.}
\label{fig:VirialParam_hex4panels_ALLeffs}
\end{figure}


\begin{figure*}
\includegraphics[width=0.88\textwidth]{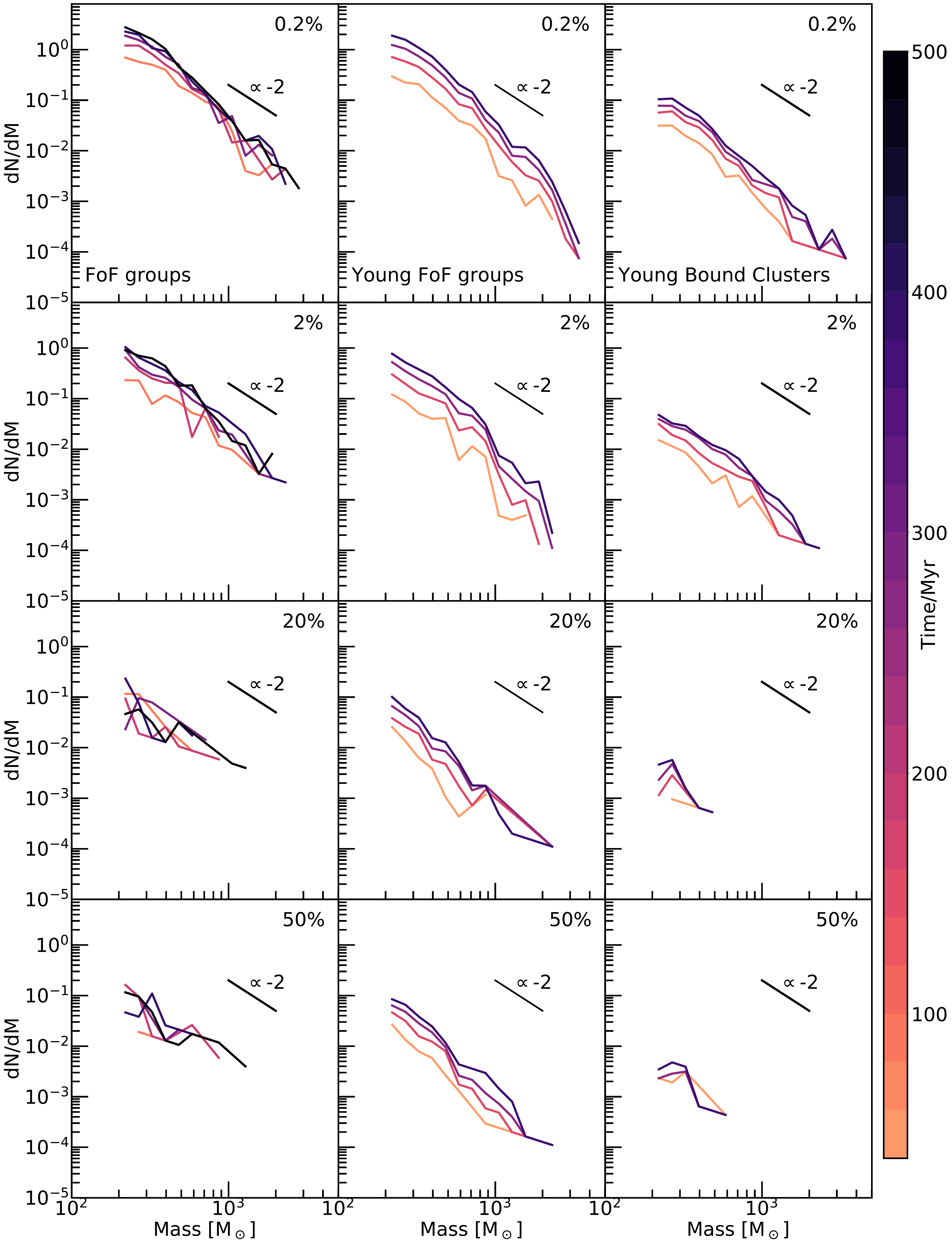}
\caption{Cluster Mass Function (CMF) of all FoF groups and clusters identified in the \ptwo, \two, \twenty\ and \fifty\ simulations. The observed slope of around -2 has been added to guide the eye. \textit{First column:} FoF groups of all ages. \textit{Second column:} Young FoF groups with ages less than 20 Myr. \textit{Third column:} Young Bound Clusters with ages less than 20 Myr. The Young Bound Clusters are thus capturing the CMF with which star clusters are born. Due to low number statistics particularly of the \twenty\ and \fifty\ runs, here we have stacked the Young FoF groups and Young Bound Clusters with ages less than 20 Myr in 20 Myr intervals. Each interval contains groups/clusters identified between 100-180 Myr, 200-280 Myr, 300-380 Myr and 400-480 Myr. All of the simulations have CMFs with slopes visually consistent with -2, however the slopes are quantified fully in Figure \ref{fig:CMF_SFR}. It is important to note that even with stacking, the low number of FoF groups clusters present in the \twenty\ and \fifty\ runs means that making any statements or ruling out any of the models based on the CMF is not possible.}
\label{fig:CMF_FoF_YoungFoF_Bound_YoungBound_alleffs}
\end{figure*}

To determine if a FoF group is a bound cluster, we perform a virial analysis by calculating the kinetic and potential energies of all stellar particles in a given FoF group. The kinetic energy is calculated using the velocities normalised to the center-of-mass motion of the FoF group. The potential energy is computed directly by calculating the potential between each pair of particles. We only consider stars and exclude gas or dark matter from this analysis. The virial parameter for each FoF group is calculated as $\alpha_{\mathrm{vir}}  = -U/2K$ where U is the sum of the potential energies and K is the sum of the kinetic energies of all stellar particles within the FoF group. An $\mathrm{\alpha}$ parameter of more or equal to one therefore denotes a bound star cluster.

Fig. \ref{fig:VirialParam_hex4panels_ALLeffs} shows the virial parameters of FoF groups with ages less than 20 Myr formed between 200 and 500 Myr in the simulation, prior to the unbinding procedure, described later in Sec. \ref{subsec:unbindingprocedure}. By plotting the FoF groups younger than 20 Myr, we capture how bound the stellar groups are shortly after their formation. 


With increasing star formation efficiency, we see a decrease in total number and a decrease of the typical virial parameter of the FoF groups. It is important to note that plotted here are simply physical associations, and therefore these groups of stars are likely to contain contaminate stars. It is however interesting to see that such a high fraction of identified FoF groups which are still likely to contain contaminating stars have such high virial parameters for the \ptwo\ and \two\ runs. Jumping from \two\ to \twenty\ however shows a steep decrease in the fraction of bound clusters. These higher star formation efficiencies result in a high fraction of FoF groups with virial parameters below 1. Some of these clusters are truly unbound but some are also contaminated with high velocity stars. The unbinding procedure explained in Section \ref{subsec:unbindingprocedure} removes these contaminate stars. One can note that the bound fraction increases again from the \twenty\ to the \fifty\ run, but with a very low number of FoF groups identified in these simulations, this is likely just low number statistics. 


\subsection{Unbinding procedure}
\label{subsec:unbindingprocedure}
As described in Section \ref{subsec:VirialAnalysis}, we calculate the overall potential and kinetic energies of the FoF groups in order to determine if they are bound. Therefore, following the FoF analysis which identifies physical associations of stars, we perform an energetic analysis of the FoF groups in order to remove contaminating stars (or determine if they are completely unbound). This is done by first sorting the stars by their distance from the centre of mass and then calculating the potential energy of every member of the FoF group in relation to the other members. Working from the outside of the FoF group inwards, a star is defined to be bound if its potential energy exceeds its kinetic energy (normalised to the bulk motion of the overall FoF group in relation to the galaxy). When a star is determined to be unbound, it is removed from the group and the potential energies of the remaining stars are updated to exclude it. This is repeated for all stars within a FoF group. When a FoF group survives this procedure, that is at least 35 members remain \citep[following the definition from][]{LadaLada2003}, the FoF group becomes classified as a bound cluster.



\subsection{Cluster mass function}
\label{subsec:CMF}
In Fig. \ref{fig:CMF_FoF_YoungFoF_Bound_YoungBound_alleffs} we show the mass function for the \ptwo, \two, \twenty\ \& \fifty\ simulations (from top to bottom) and FoF groups, young FoF groups and young bound clusters (from left to right). The mass functions are plotted at different times as indicated by the colour bar. For comparison we show a typical mass function with a power-law slope of $-2$ \citep[e.g.][]{Larsen09, Gieles09, Zhang99, Vansevicius09,ZwartReview2010} in each panel. Such a slope is favoured by observations. We find a very low number of clusters in the \twenty\ and \fifty\ runs. We therefore decide to stack the young FoF groups together, as well as the bound clusters. The reason for this is simply to increase the number of clusters. Without stacking, for the \twenty\ and \fifty\ runs, identifying a slope becomes difficult. The second and third columns therefore show the stacked cluster mass function (CMF) of young FoF groups and young bound clusters, respectively, with average ages of less than 20 Myr. The third column therefore captures the CMF in which bound clusters form.

\begin{figure}
\includegraphics[width=\columnwidth]{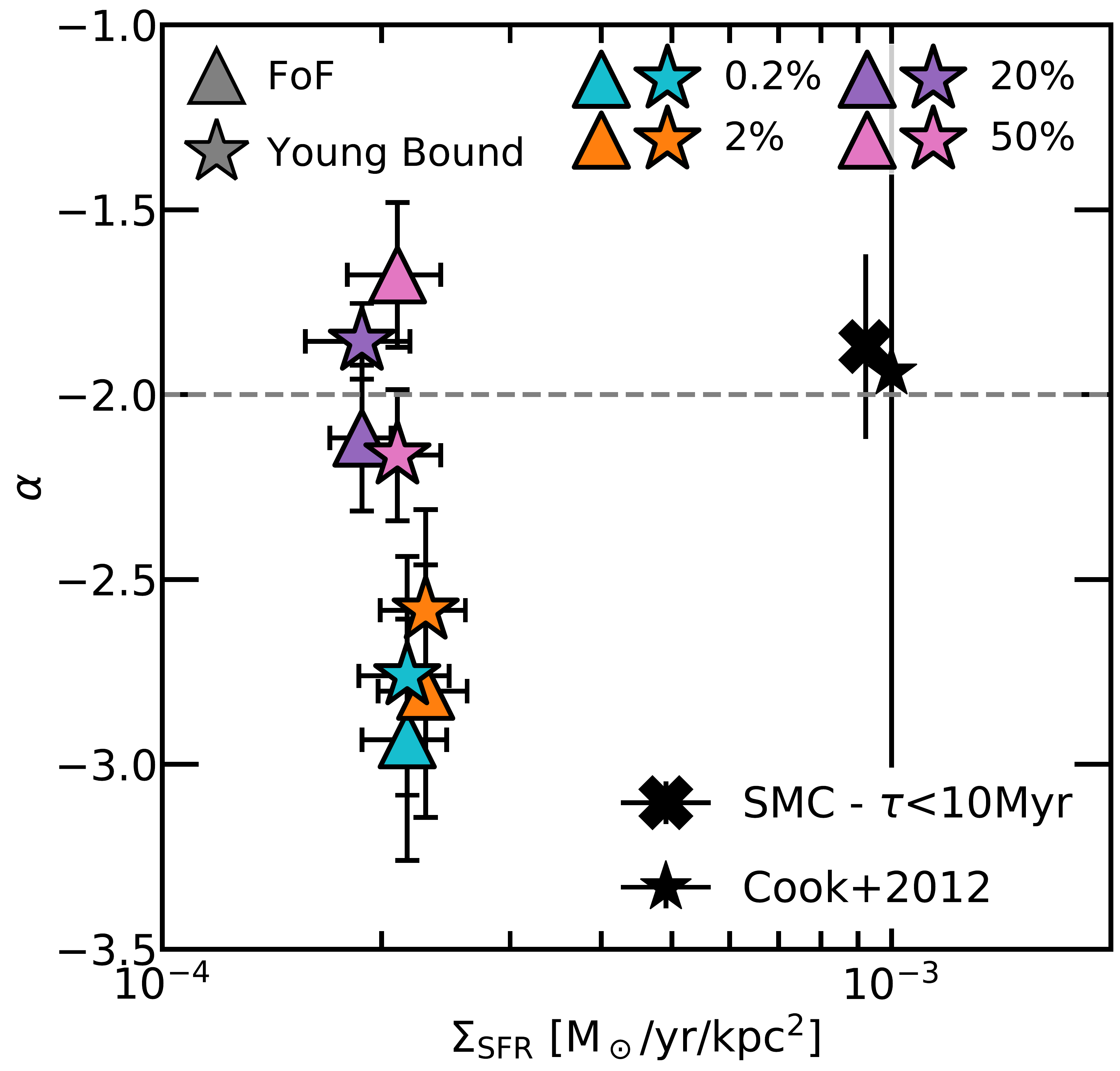}
\caption{The slope, $\alpha$ of the cluster mass function $\mathrm{(dN/dM)}$ as a function of the star formation rate surface density. Both quantities are averaged between 300 to 500 Myr. The slope is shown for FoF groups (triangles) and bound clusters younger than 10 Myr in age (stars) with the colours representing the different simulations. Vertical error bars here show the standard deviation from the fit of $\alpha$ and horizontal error bars show the standard deviation of the mean \Ssfr\ averaged over 300-500 Myr. We see that lower \sfe\ results in steeper slopes of the cluster mass function.}
\label{fig:CMF_SFR}
\end{figure}

The average slopes from Fig. \ref{fig:CMF_FoF_YoungFoF_Bound_YoungBound_alleffs} are summarised in Fig. \ref{fig:CMF_SFR}, showing the slope $\mathrm{\alpha}$ of the CMF plotted as a function of the star formation rate surface density \Ssfr. The \Ssfr\ is calculated within a circle of 1 kpc placed over the face-on disk and 500 pc above and below the plane of the disk. This encompasses >99\% of star formation in the disk for all simulations. For $\mathrm{\alpha}$, we see that increasing the star formation efficiency parameter results in shallower slope. We find that both the slope of the FoF groups as well as the bound clusters are both in agreement with $-2$, implying that this slope is universal and not just for bound structures. This ties into the hierarchical distribution of star formation where clouds, stellar associations as well as bound star clusters all broadly follow a slope of $-2$ \citep{Elmegreen2011}. As mentioned, a slope of approximately $-2$ is the broadly accepted value of the cluster mass function of observed star clusters, however there is some variation in the literature. \citet{Adamo2020b} find a slope of between -1.5 and -2 for a population of star clusters within the Hubble imaging Probe of Extreme Environments and Clusters (HiPEEC) survey. \citet{Chandar2017} find a slope of $-2$ is consistent for a range of masses of objects, and it is worth noting that they also split their clusters by age and find no change in the slope, only in the normalisation of the CMF. From our simulations, young bound clusters formed in the \twenty\ and \fifty\ runs have slopes close to $-2$, therefore from the CMF alone, this supports a higher value for \sfe. This, however, is based on a small number of clusters. Something to consider is that observations are naturally biased towards more massive clusters. However, when correctly taking incompleteness into account, there should not be a large effect on the slope of the CMF.



\subsection{Cluster formation efficiency}
\label{subsec:CFE}


\begin{figure}
\includegraphics[width=\columnwidth]{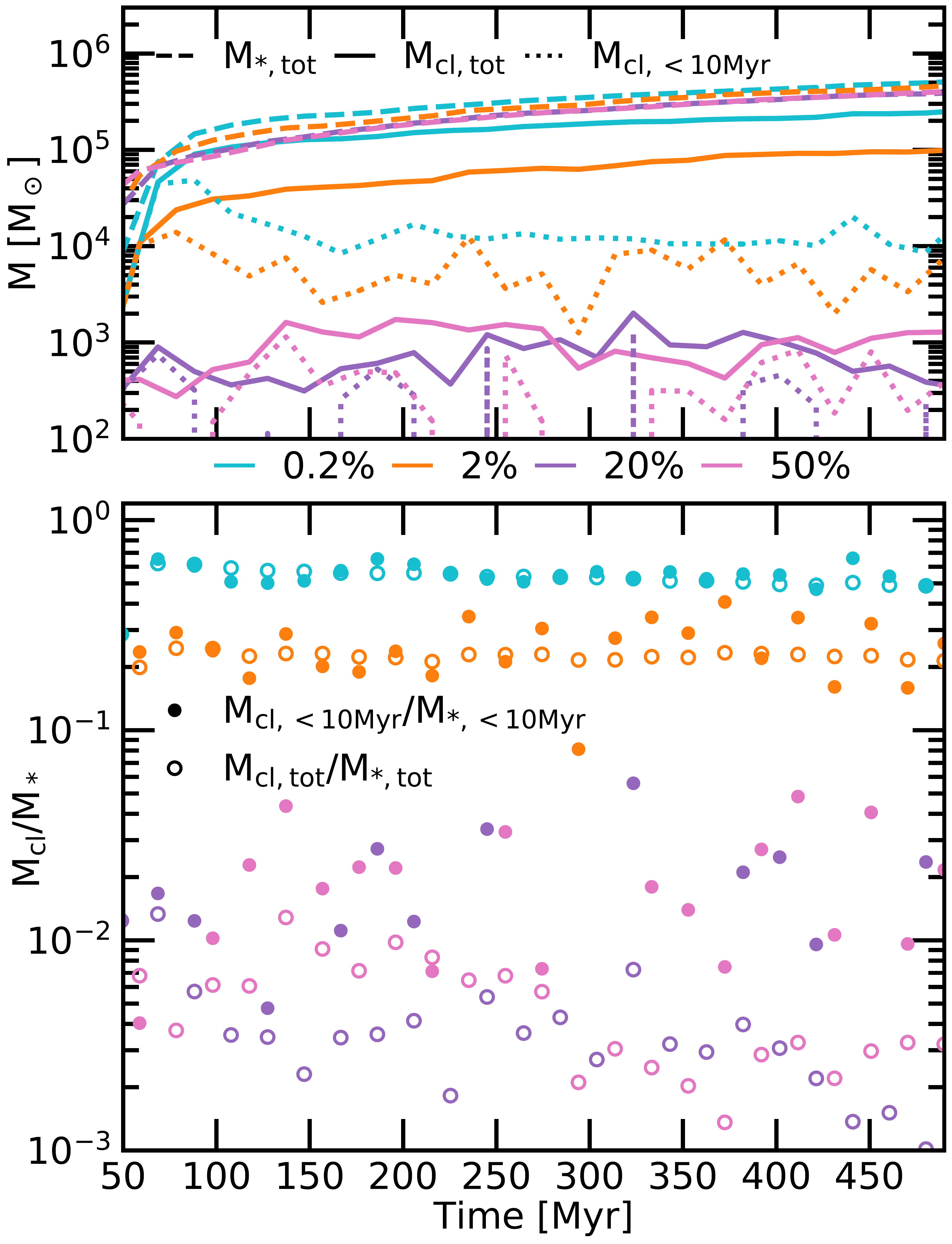}
\caption{\textit{Top panel:} Time evolution of the total mass of newly formed stars (dashed line), stellar mass in bound clusters (solid line) and stellar mass in young bound clusters with ages less than 10 Myr (dotted line). At low star formation efficiencies, \ptwo\ and \two, the total mass in clusters (solid line) follows the stellar mass build up as the clusters are not disrupted. For the higher efficiency runs (\twenty\, and \fifty) the mass in bound clusters at all ages stays constant, indicating the disruption of bound clusters. \textit{Bottom panel:} Time evolution of the mass in clusters as a fraction of the total stellar mass. Filled circles show the cluster formation efficiency, defined as the fraction of young (<10 Myr) stellar mass in bound clusters. Open circles represent the fraction of mass in bound clusters of all ages. We do not identify any young clusters in 46 per cent and 17 per cent of the time for the \twenty\ and \fifty\ runs respectively.}
\label{fig:CFE_Bound_ALLeffs}
\end{figure}

\begin{figure*}
\includegraphics[width=0.95\textwidth]{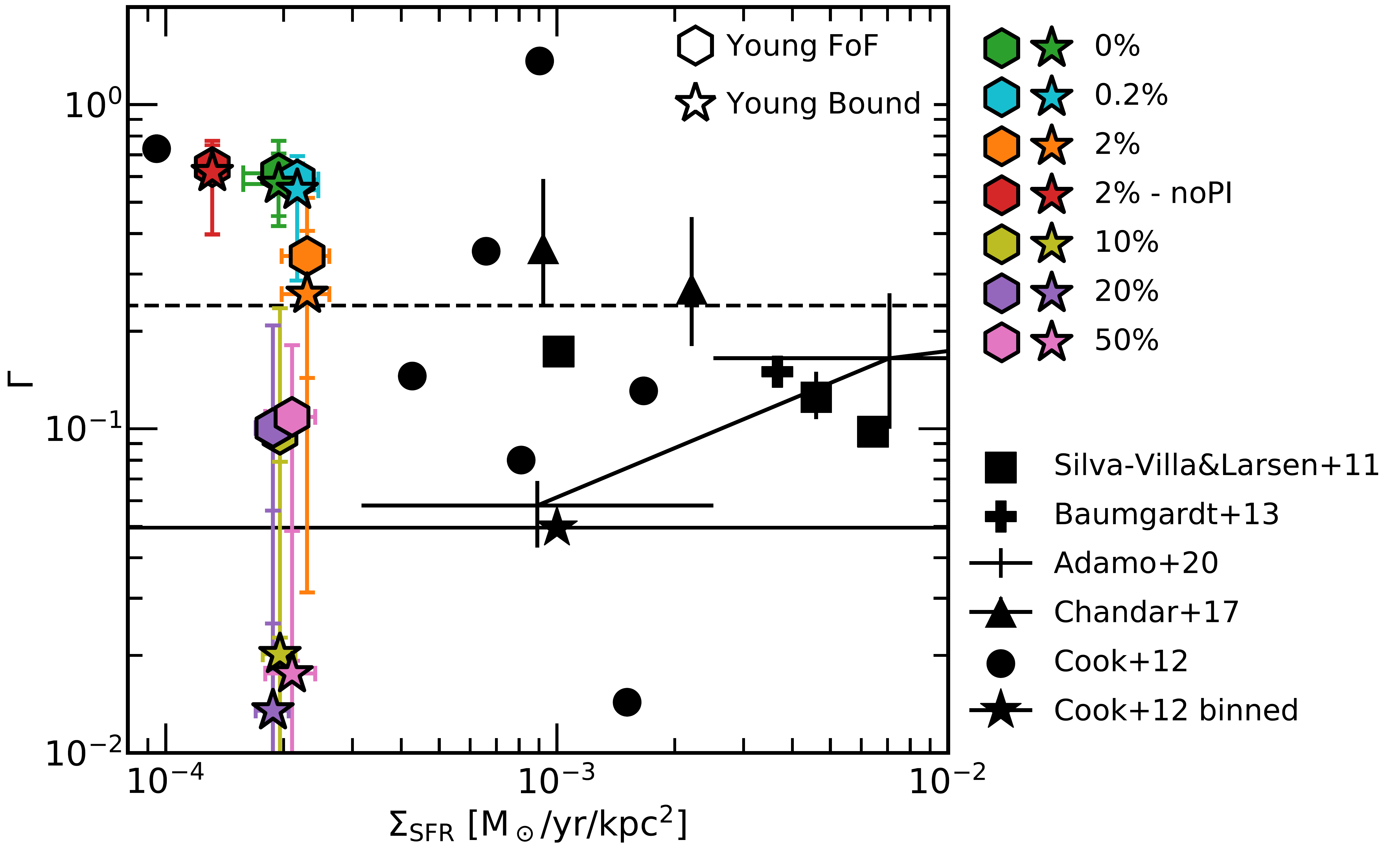}
\caption{Average cluster formation efficiency $\mathrm{\Gamma}$ for each model for young FoF groups (crosses) and young bound clusters (circles) as a function of average star formation rate surface density, $\mathrm{\Sigma_{SFR}}$. Averages are calculated by taking the cluster formation efficiency of FoF groups or bound clusters with ages younger than 20 Myr at 20 Myr intervals in the simulations between 300 and 500 Myr. We show observational data of NGC6946, NGC45, and NGC7793 \citep{Silva-VillaLarsen11}, LMC \citep{Baumgardt13} as well as the median of a collection of data compiled in a review by \citet{Adamo2020} for young clusters with ages younger than 10 Myr. We also show data for the SMC and LMC from \citet{Chandar2017}, who find a constant $\Gamma \sim$ 24 per cent, irrespective of \Ssfr\ for clusters with ages 1-10 Myr (dashed horizonal line). Individual observations of young (<10 Myr) clusters in nearby dwarf galaxies from \citet{2012ApJ...751..100C} hint at a negative trend between $\Gamma$ and \Ssfr. The median of the data from \citet{2012ApJ...751..100C} binned between 10$^{-4.5}$ and 10$^{-2}$ in \Ssfr\ is also shown.}
\label{fig:Gamma_SFR}
\end{figure*}


In the top panel of Fig. \ref{fig:CFE_Bound_ALLeffs} we show the time evolution of the total mass of stars formed (dashed lines) as well as the total mass of stars in bound clusters (solid lines) and the mass in young bound clusters with ages younger than 10 Myr (dotted lines). For the low star formation efficiency runs (blue and orange), we see a similar trend. The mass in bound clusters (all ages, solid line) increases steadily with the mass of stars formed (dashed line), indicating we have a constant fraction of stars in bound clusters at any given time. This is shown quantitatively in the bottom panel by the open circles which show the fraction of mass in bound clusters of all ages. We observe a roughly constant fraction of $\sim 0.5$ for the \ptwo\ and $\sim 0.2$ for the \two\ run. The cluster formation efficiency (CFE, $\Gamma$) is usually defined as the fraction of the mass in young stars that have formed in clusters \citep[see e.g.][]{1997ApJ...480..235E,Kruijssen2012}, which is shown in the bottom panel of Fig. \ref{fig:CFE_Bound_ALLeffs} with filled circles. For the low efficiency runs, this value remains roughly constant (solid blue and orange circles, bottom panel of Fig. \ref{fig:CFE_Bound_ALLeffs}) indicating that newly formed clusters, which are all very bound (see Fig.\ref{fig:VirialParam_hex4panels_ALLeffs}), are not disrupted. For the high efficiency runs \twenty\ and \fifty, the situation is quite different. Both the mass in bound clusters at all ages (pink and purple solid lines, top panel of Fig. \ref{fig:CFE_Bound_ALLeffs}) and the mass of newly formed clusters (pink and purple dotted lines) stay constant. The total mass in young bound clusters remains constant, whilst the stellar mass steadily increases. We see that the fraction mass formed in young clusters (filled pink and purple circles, bottom panel of Fig. \ref{fig:CFE_Bound_ALLeffs}) stays roughly constant, whilst the overall mass in clusters decreases (open pink and purple circles). This is clear evidence for the disruption of bound clusters in these simulations. We also find in 46 per cent and 17 per cent of snaps in the \twenty\ and \fifty\ runs we do not identify any clusters at all. Qualitatively, we have seen this behaviour already Fig. \ref{fig:ALLeffs_6panels_stars_snap400} where the stellar distribution is more smooth for the higher SFE runs. Clusters with a lower virial parameter are less bound (see Fig. \ref{fig:VirialParam_hex4panels_ALLeffs}) and therefore more susceptible to internal and external disruption processes.

We take the average CFE between 300 and 500 Myr from Fig. \ref{fig:CFE_Bound_ALLeffs} and show this as a function of the corresponding average star formation rate surface density \Ssfr\ in Fig. \ref{fig:Gamma_SFR} for all simulations presented in this paper. \Ssfr\ is calculated as described in \ref{subsec:CMF} and the CMF is calculated over the same area, which is a cylinder with radius of 1 kpc and a height of 500 pc. We show the young FoF groups (hexagons) as well as young bound clusters (stars). For the lowest star formation efficiencies, \zero\ (green) and \ptwo\ (blue), more than $\sim 50$ \% of stars are born in FoF groups or bound clusters (most FoF groups are bound, see Fig. \ref{fig:VirialParam_hex4panels_ALLeffs}). In contrast, the \ten, \twenty\ and \fifty\ runs (lime green, pink and purple) have averages of only $\sim 1 - 2$ per cent for young bound clusters and $\sim 10$ per cent for young FoF groups. For these high efficiency simulations, most young FoF groups are not bound (see Fig. \ref{fig:VirialParam_hex4panels_ALLeffs}). 
The observational data in Fig. \ref{fig:Gamma_SFR} show the median of data compiled in a recent review by \citet{Adamo2020}, which reveals a positive correlation between $\Gamma$ and \Ssfr\ for young clusters (<10 Myr). It is important to note however that not all observational literature supports this relation. \citet{Chandar2017} comment on the fact that for data at high \Ssfr, $\Gamma$ is preferentially estimated on short time scales (e.g. 1-10 Myr), whilst data at low \Ssfr\ is estimated over a longer time scale, up to $100$ Myr. Accounting for this, when only considering newly formed stars over an age range of $1-10$ Myr, they find a constant $\Gamma$ of around 24 per cent, irrespective of \Ssfr, shown as the dashed horizontal line. The observational results have made varying assumptions on the definition of clusters, details of which are discussed in a recent review by \citet{Adamo2020}. We also include data for young (<10 Myr) clusters from \citet{2012ApJ...751..100C} who look at local dwarf galaxies. Their data hint at a negative correlation between $\Gamma$ and \Ssfr, which is in slight tension with \citet{Adamo2020}. However, as is discussed in both these studies, cluster formation is highly stochastic and heavily dependent on the evolutionary phase of each individual galaxy. The majority of observational data suggest a positive correlation between $\Gamma$ and \Ssfr, however the significant scatter means that none of our simulations with differing values of \sfe\ can necessarily be ruled out by $\Gamma$ alone.

Linking back to the global properties discussed in Sec. \ref{subsec:GlobalProperties}, it is interesting to remind ourselves that with these different cluster formation efficiencies leading to very different clustering properties, we do not see an effect on the outflow properties. 







\subsection{Cluster ages}
\label{subsec:AgeSpreads}
Observations of the age spreads of star clusters are challenging as projection effects can introduce contamination from older stars. From a variety of objects studied in the literature, an appropriate upper limit for the expected age spread would be approximately 5 Myr \citep{Longmore2014}. We look at the age spreads of the bound clusters in the simulations, which are shown in Table \ref{table:GalaxyProperties}. In the \ptwo\ and \two\ models where we also have the most clusters, the majority of clusters have age spreads of less than $5$ Myr, but some have spreads of up to $20-30$ Myr. For the \twenty\ and \fifty\ models, we have less bound clusters overall to examine, but these clusters have wider age spreads. This provides some support against these models with higher star formation efficiencies as the star formation histories of the individual clusters are more extended.



\subsection{Cluster sizes}
\label{subsec:ClusterSizes}

\begin{figure}
\includegraphics[width= \columnwidth]{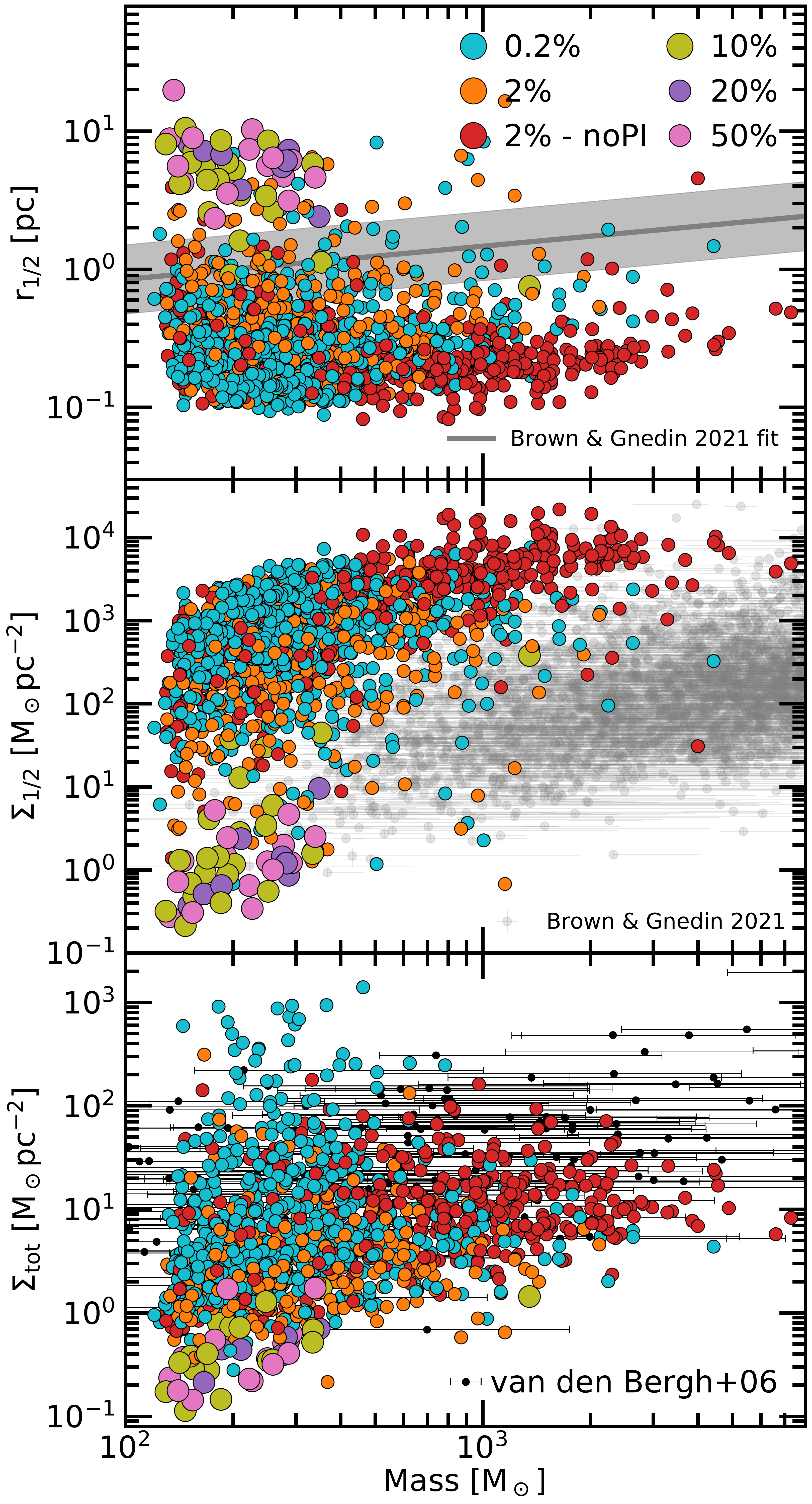}
\caption{\textit{Top panel:} Half-mass radius of bound clusters as a function of cluster mass for \ptwo, \two, \nophoto, \ten, \twenty\ and \fifty. The sizes of the symbols for the \ten, \twenty\ and \fifty\ runs have been artificially increased for visual clarity. The solid line indicates the best fit relation with scatter from LEGUS observations of 31 galaxies presented in \citet{2021arXiv210612420B}. \textit{Middle panel:} Half-mass surface density of bound clusters. Observations from \citet{2021arXiv210612420B} are shown for comparison. \textit{Bottom panel:} Total surface density of bound clusters as a function of their total mass. Observations from \citet{2006AJ....131.1559V} are shown for comparison.}
\label{fig:MassDensity3Plots_BoundClusters_ALLEFFS_snaps200500}
\end{figure}

In Fig. \ref{fig:MassDensity3Plots_BoundClusters_ALLEFFS_snaps200500} we show, from top to bottom, the half mass radius $\mathrm{r_{1/2}}$, half mass surface density $\Sigma_{1/2}$ and total surface density $\Sigma_{\mathrm{tot}}$, defined as the density of the region containing 90 per cent of the cluster mass, as a function of total cluster mass for the bound clusters identified in the simulations.  In the \twenty\ and \fifty\ runs, we see that the clusters are more extended with half mass radii of around 3-10 pc, in comparison to the \ptwo\ and \two\ runs which have half mass radii mostly an order of magnitude lower, around 0.1-1 pc. The \ten\ run shows clusters with similar sizes to the \twenty\ and \fifty\ with a few more compact clusters with $\mathrm{r_{1/2}}$ of around 1 pc. The \ten, \twenty\ and \fifty\ runs also have lower surface densities consistent with their lower virial parameters (see Fig. \ref{fig:VirialParam_hex4panels_ALLeffs}). For comparison we show the recently published mass-size relation from \citet{2021arXiv210612420B}. It appears that the simulated clusters are either too small for low star formation efficiencies or too large for high star formation efficiencies.  The half-mass surface densities of the low star formation efficiency simulations seem to be lower than observed (middle panel of Fig. \ref{fig:MassDensity3Plots_BoundClusters_ALLEFFS_snaps200500}), while the total surface densities (bottom panel) seem consistent with observations. This indicates that clusters produced in simulations with low \sfe\ below 10$^3$ \solarmass\ are too compact compared to observations. The situation improves for the few simulated clusters at higher mass \citep[see also][for simulated clusters in a starburst environment]{Lahen2020}. The clusters from the \ten, \twenty\ and \fifty\ runs are likely too diffuse. We discussed previously in Section \ref{subsec:CFE} that there is cluster disruption in the higher star formation efficiency models. Observationally, it is expected that star clusters disrupt, and capturing these disruption properties appears to be a challenge of these galaxy formation simulations. If we do observe cluster disruption in these runs, it is likely due to the fact that the surface densities were too low at formation. 
It is worth noting as a caveat however that the cluster sizes and properties are likely heavily affected by the fact that the interactions between the stars are softened, reducing dynamical effects such as two-body relaxation. Discussion of this as well as proposed solutions are discussed in Sec. \ref{subsec:scdisruption}.

\section{Discussion}
\label{sec:Discussion}

\subsection{Star formation and outflow rates}

For the same set of physical processes included in the simulations, the choice of \sfe\ significantly changes the properties of the forming star clusters but has little impact on the global galaxy properties such as SFR or mass loading. This is in agreement with previous studies (see Sec. \ref{sec:Introduction}). When excluding photo-ionisation, we form significantly more clusters. However, the peaks in the ambient supernova densities only slightly change. After the initial starburst, mass loading is increased by approximately a factor of two, which is in agreement with the simulations shown in \citet{2021MNRAS.506.3882S}, indicating that stronger stellar clustering in their model without HII regions leads to higher mass loading factors, also by a factor of approximately two. Their modelling of heating/cooling, star formation, and HII regions is similar to the one used here. \citet{2021MNRAS.506.3882S} also find that the exclusion of photo-ionisation increases the maximum ambient densities at which SN explode quite substantially as well as broadening the distribution of ambient densities, in agreement with \citet{Hu2017} as well as our findings.

Over the entire simulation, a substantial part of the regulation of star formation is done by HII regions, which also result in smoother star formation histories. This conclusion seems very robust as it has also been put forward by earlier studies  with varying setups and simulation codes \citep[see e.g.][]{2011MNRAS.417..950H,Peters2017,2017ApJ...841...82B,2018MNRAS.480..800H,2019MNRAS.482.4062H,2020MNRAS.491.2088K,2020MNRAS.491.3702H,2021MNRAS.506.3882S,2021MNRAS.504.1039R}.  The formation of HII regions also reduces clustering and cluster growth \citep[e.g.][]{2018MNRAS.477.5001G,2021MNRAS.506.3882S,2021MNRAS.504.1039R}. We note however that beyond the initial starburst, the \nophoto\ simulation has a lower star formation rate.

\subsection{Star cluster disruption}
\label{subsec:scdisruption}
As this study shows, the observed star cluster mass function can be successfully reproduced with high-resolution galaxy evolution simulations. Also the fraction of stars forming in clusters can be controlled by varying the star formation efficiency. However, another fundamental star cluster property cannot be modelled yet, which is the relatively rapid destruction of star clusters after formation  \citep[see e.g.][]{ZwartReview2010,2019ARA&A..57..227K}. In our study, only simulations with very high \sfe\ values show signs of cluster disruption. These clusters however, are too diffuse compared to observations and can therefore easily be dispersed by tides, which are naturally included in the simulations. The simulated clusters presented in \citet{2017MNRAS.464.3580D} show similar properties. No high resolution galaxy formation simulation to date produces star cluster properties which are all consistent with observations. The reason could be the still limited capability to properly resolve internal star cluster properties in galaxy evolution simulations. Important caveats here might be: 
\begin{itemize}
\item limitations for resolving the internal star-forming cloud structure and dynamics on sub-parsec scales
\item inability of the \sfe\ based star formation sub-grid model to capture the accurate distribution and timing of star formation and stellar feedback
\item inability of the feedback model to accurately capture gas expulsion from forming star clusters \citep[see e.g.][]{2006MNRAS.369L...9B}
\item missing feedback channels like stellar winds 
\item limited capabilities of the typically used second order integrators to follow important relaxation effects on cluster scales. 
\end{itemize} 

A possible solution might require a turning away from \citet{1959ApJ...129..243S}-type star formation models. Additionally, the interactions between the stars themselves are softened within the simulations, reducing the two-body interactions within the clusters. These interactions are essential in both the formation as well as the evolutionary fate of the clusters. \citet{Gieles10} show that by 10 Myr, two-body relaxation has had a strong effect on the evolution of globular clusters. Accurately modelling these interactions could therefore be achieved by the use of higher order forward integration schemes \citep[see e.g][]{2021MNRAS.502.5546R}, which allow for higher dynamical fidelity in dense stellar systems and a straight forward coupling with hydrodynamics.

\section{Conclusions}
\label{sec:Conclusions}

We present high-resolution (sub-parsec, 4 \solarmass) simulations of the evolution of dwarf galaxies. The simulations include non-equilibrium heating and cooling processes and chemistry, an interstellar radiation field varying in space and time, star formation, a simple model for HII regions as well as supernova explosions from individual massive stars.  We explore the impact of assumed star formation efficiencies, \sfe\, per free-fall time for a \citet{1959ApJ...129..243S}-type formation model on the resulting star formation and outflow rates and the star cluster properties.  We find the following results: 

\begin{enumerate} 
    \item Star formation rates and outflow rates are independent of \sfe\ for the investigated range of \sfe\ = 0.02 - 0.5 (\ptwo, \two, \ten, \twenty\ and \fifty) and a model with instantaneous star formation at a high density threshold (\zero). The test model without HII regions (\nophoto) has a similar star formation rate, but a slightly higher outflow rate resulting in a slightly increased mass loading. 
    \item All simulations form star clusters with power law mass functions similar to observations. With increasing \sfe, the slope $\alpha$ increases from -3 to -2. The normalisation of the cluster mass function, i.e. the mass of the most massive cluster formed, decreases with increasing \sfe. At higher \sfe, clusters are less likely to form as well as survive due to the fact that the stars are formed at lower ambient densities.
    \item The clusters become less bound and the cluster formation efficiencies decrease from $\Gamma \sim 0.6$ to $\Gamma \sim 0.1$ with increasing \sfe. The physical reason for this is due to the fact that changing the \sfe\ controls the densities at which the stars form, as shown in Sec. \ref{subsec:ambientdensitySNII}. Low star formation efficiencies mean more star formation at higher densities resulting in massive, compact, bound clusters. The low formation star efficiency (\sfe\ = 0.02, \ptwo)  and the instantaneous formation model (\zero) are inconsistent with all available cluster formation efficiency observations.
    \item None of the models seem to match observed cluster sizes. Clusters in simulations with high efficiencies \sfe\ $\gtrsim$ 0.2 are too diffuse. While they shown signs for cluster disruption these models are disfavoured as no internal rapid cluster evolution process can make cluster more compact. Clusters in low \sfe\ simulations are too compact and do not disrupt. A more accurate modelling of internal evolutionary processes might be able to alleviate this problem.
\end{enumerate}

The failure of the current highest resolution galaxy evolution models to capture all fundamental star cluster properties poses a challenge for all future numerical studies on galactic star cluster populations. 

\section*{Data Availability}
The data will be made available based on reasonable request to the corresponding author.

\section*{Acknowledgements}
We thank the referee, Nate Bastian, for valuable comments and suggestions on the submitted manuscript. The authors also thank Rebekka Bieri, Francesca Fragkoudi, Matthew Smith, and Vadim Semenov for insightful discussions, as well as Miha Cernetic for technical support. UPS is supported by the Simons Foundation through a Flatiron Research Fellowship at the Center for Computational Astrophysics. The Flatiron Institute is supported by the Simons Foundation. PHJ and DI acknowledge support by the European Research Council via ERC Consolidator Grant KETJU (no. 818930).
We acknowledge the computing time provided by the Leibniz Rechenzentrum (LRZ) of the Bayrische Akademie der Wissenschaften on the machine SuperMUC-NG (pn72bu). This research was supported by the Excellence Cluster ORIGINS which is funded by the Deutsche Forschungsgemeinschaft (DFG, German Research Foundation) under Germany's Excellence Strategy – EXC-2094 – 390783311. We thank the super computing resources at the LRZ in Garching for using an energy mix that is to $100$ per cent comprised out of renewable energy resources (e.g. \url{https://www.top500.org/news/germanys-most-powerful-supercomputer-comes-online/}, \\\url{https://www.lrz.de/wir/green-it_en/}). This work made use of the software packages \textsc{NumPy} \citep[][]{Harris20}, \textsc{SciPy} \citep[][]{Virtanen20}, \textsc{AstroPy} \citep[][]{Price-Whelan18}, \textsc{Jupyter} \citep[][]{Kluyver16}, \textsc{Matplotlib} \citep[][]{Hunter07} and \textsc{Pygad}\footnote{\url{https://bitbucket.org/broett/pygad/src/master/}} \citep{2020MNRAS.496..152R}.




\bibliography{SCpaper_bib} 

\bsp	
\label{lastpage}
\end{document}